\documentclass[%
 reprint,
 superscriptaddress,
 amsmath,amssymb,
 aps,
 prl,
]{revtex4-1}

\usepackage{graphicx}
\usepackage{bm}

\begin{document}

\title{Thermal Conductivity and Electrical Resistivity of Solid Iron at Earth's Core Conditions from First Principles}

\author{Junqing Xu}
\affiliation{Department of Earth and Environmental Sciences, LMU Munich, Theresienstrasse 41, 80333 Munich, Germany}
\author{Peng Zhang}
\affiliation{School of Science, Xi'an Jiaotong University, Xi'an, Shaanxi, 710049, China}
\author{K. Haule}
\affiliation{Department of Physics, Rutgers University, Piscataway, New Jersey 08854, USA}
\author{Jan Minar}
\affiliation{University of West Bohemia, New Technologies - Research Centre, Pilsen, Czech Republic}
\author{Sebastian Wimmer}
\affiliation{Department Chemie,  Physikalische Chemie, University of Munich, D-81377 Munich, Germany}
\author{Hubert Ebert}
\affiliation{Department Chemie,  Physikalische Chemie, University of Munich, D-81377 Munich, Germany}
\author{R. E. Cohen}
\email{rcohen@carnegiescience.edu}
\affiliation{Department of Earth and Environmental Sciences, LMU Munich, Theresienstrasse 41, 80333 Munich, Germany}
\affiliation{Extreme Materials Initiative, Geophysical Laboratory, Carnegie Institution for Science, Washington, D.C. 20015-1305, USA}

\date{Received 26 March 2018}

\begin{abstract}
We compute the thermal conductivity and electrical resistivity of solid hcp Fe to pressures and temperatures of Earth's core.
We find significant contributions from electron-electron scattering, usually neglected at high temperatures in transition metals.
Our calculations show a quasilinear relation between the electrical resistivity and temperature for hcp Fe at extreme high pressures.
We obtain thermal and electrical conductivities that are consistent with experiments considering reasonable error.
The predicted thermal conductivity is reduced from previous estimates that neglect electron-electron scattering.
Our estimated thermal conductivity for the outer core is 77$\pm$10 W\,m$^{-1}$\,K$^{-1}$, and is consistent with a geodynamo driven by thermal convection.
\end{abstract}

\maketitle

The thermal conductivity of iron (Fe) and its alloys at Earth's core conditions is of central importance to understanding
the thermal evolution of Earth's core and the energetics of the geomagnetic field
\cite{labrosse2003thermal,nimmo2015treatise,stacey2007revised}.
A wide range of values for the thermal conductivity at core conditions has been predicted
\cite{stacey2007revised,sha2011first,de2012electrical,pozzo2012thermal,zhang2015effects,secco2017thermal}.
Previously, the thermal conductivity of iron at extreme conditions has been obtained from the electrical resistivity
by applying the Wiedemann-Franz law: $\kappa = L T \sigma$,
where $\kappa$ and $\sigma$ are the thermal and electrical conductivities, respectively, $\sigma$ is the inverse of electrical resistivity $\rho$,
and $L$ is the conventional Lorenz number $L_0$ (2.44$\times$10$^{-8}$~W$\Omega$K$^{-2}$)
\cite{keeler1969electrical,gomi2013high,ohta2016experimental}.
The Wiedemann-Franz law has generally not been verified for any material at extreme conditions. It can be derived under approximations  \cite{jonson1980mott} that would not apply under the high temperature of Earth's core.  Direct measurements of thermal conductivity at conditions close to
Earth's core conditions gave low values (e.g., 46 W m$^{-1}$ K$^{-1}$) \cite{konopkova2016direct}
that would support the conventional thermal dynamo picture and are consistent with a geodynamo operating via thermal convection through Earth history. However, the thermal conductivity measurements and electrical resistivity measurements \cite{ohta2016experimental} are inconsistent, requiring extreme violations of the Wiedemann-Franz law.
High values of thermal conductivity (220 W m$^{-1}$ K$^{-1}$) predicted by first-principles molecular dynamics (FPMD) with the Kubo-Greenwood formula
within density functional theory (DFT) \cite{pozzo2012thermal} are inconsistent with thermal convection of the core, requiring a different mechanism
\cite{o2016powering,badro2016early}.
In addition to the relationship to heat transport, the electrical resistivity of iron and its alloys at Earth's core conditions is an important quantity for the geodynamo in itself, since a higher resistivity increases the dynamo dissipation.

We computed both the electron-phonon ($e$-ph) and electron-electron ($e$-$e$) scattering contributions to electrical and thermal conductivity in solid hcp iron.
For each contribution, we have used two methods that have complementary approximations.
First, we computed the $e$-ph contribution using the density functional perturbation theory (DFPT) and the inelastic Boltzmann transport equation \cite{allen1978new} within {\sc ABINIT}
\cite{gonze2016,supplemental}.
Everywhere below, where we say ``Boltzmann theory" we refer to electron-phonon scattering computed using the DFPT and Boltzmann transport theory.
At high temperatures, the mean free path $l$ of electron due to $e$-ph scattering becomes comparable to the lattice constant
so that resistivity saturation may become important \cite{gomi2013high}.
The Boltzmann theory does not include saturation effects.
We estimate such effects by applying the parallel resistor formula \cite{wiesmann1977simple},
whose reliability has been verified theoretically \cite{allen1981infrared} and numerically \cite{schiller1982influence,werman2017non}:
\begin{eqnarray}
\frac{1}{\rho_{e-ph}} = \frac{1}{\rho_{sat}} + \frac{1}{\rho_{B}}, \label{eq:sat}
\end{eqnarray}
where $\rho_{B}$ is from the Boltzmann theory. $\rho_{sat}\,=\,\rho_{B} l_{B} / a$,
where $a$ is the lattice constant and $l_{B}$ is the mean free path, i.e., the product of the relaxation time and Fermi velocity.

Second, we computed the $e$-ph contribution using the Korringa-Kohn-Rostoker (KKR) method with the coherent potential approximation (CPA)
to model thermal lattice vibrations \cite{ebert2015calculating} within the {\sc SPRKKR} code \cite{ebert2011calculating}.
The scattering by phonons is computed from scattering by atomic displacements.
Unlike the DFPT computations above, the KKR-CPA naturally includes resistivity saturation effects,
which have been discussed as having important implications to transports properties in Earth's core \cite{gomi2013high}.
We calculate the resistivity of hcp Fe at high temperatures using the KKR-CPA and the Kubo-Greenwood formula
and find that the slope decreases with the temperature (Fig. \ref{rho_v47.8}) consistent with saturation effects.
The theoretical results by the method above using the DFPT + Eq. (\ref{eq:sat}) and the KKR-CPA are
in good agreement with each other in a wide range of pressure and temperature (Fig. 1 and 4).
However, the neglect of local environmental effects in the single-site KKR-CPA may lead to errors of transport properties.\par

\begin{figure}[t]
\includegraphics[width=1.0\columnwidth]{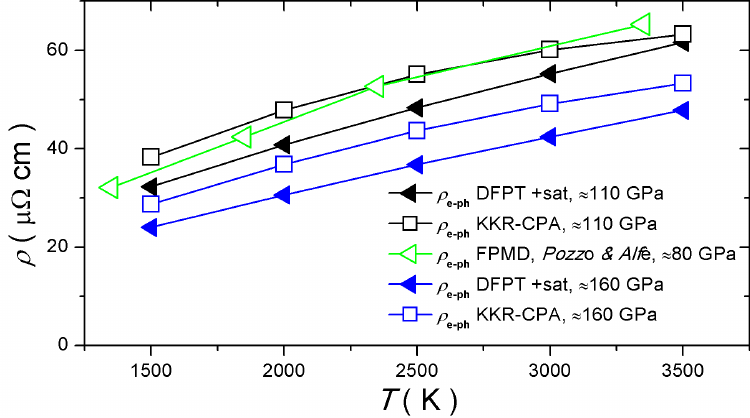}
\caption{\label{rho_DFPTvsKKR} Calculated resistivity of hcp Fe at atomic volume 57.9 and 53.3 bohr$^3$, corresponding to about 110 and 160 GPa at 2500 K \cite{dewaele2006quasihydrostatic,sha2010first}.
$\rho_{e-ph}$ is the electron-phonon contribution of resistivity.
DFPT represents calculating $\rho_{e-ph}$ using the DFPT + inelastic Boltzmann theory.
"+ sat" includes resistivity saturation effects for the $e$-ph scattering using Eq. (\ref{eq:sat}).
"KKR-CPA" represents using the KKR-CPA method with the Kubo-Greenwood formula.
FPMD is first-principles molecular dynamics with the Kubo-Greenwood formula.
Both the KKR-CPA and the FPMD have naturally included resistivity saturation effects.}
\end{figure}

We compared our thermal conductivity results with previous FPMD computations\cite{pozzo2012thermal,de2012electrical},
and find that they agree well with those in Ref. \cite{de2012electrical}, but less so with those in Ref. \cite{pozzo2012thermal}.
Within our methods, it seems to be easier to achieve good convergence than with FPMD.
In practice, parameters in FPMD simulations, especially the cell size and number of $k$ points, are difficult to be converged for high-density metals.
For instance, energy levels are discrete, and due to the finite supercell size in FPMD, the low-frequency part of optical electrical and thermal conductivity and their dc values will depend on the choice of the width of the broadening function. We do not have such issues in our methods.

We computed the $e$-$e$ scattering contribution to electrical resistivity and thermal conductivity using density functional theory + dynamical mean field theory (DFT+DMFT) \cite{georges1996dynamical,kotliar2006electronic} with the continuous time quantum Monte Carlo impurity solver \cite{haule2007quantum,gull2011continuous} and the Kubo-Greenwood formula with the {\sc EDMFTF} code \cite{haule2010dynamical,haule2015free}.
We use the method of Refs. \cite{zhang2015effects,zhang2016retraction}, but correcting the factor of 2 error there due to the neglect of the two spin channels.
The $e$-$e$ scattering is additionally studied using the KKR-DMFT \cite{minar2005multiple} with the spin-polarized $T$ matrix + fluctuation exchange impurity solver \cite{pourovskii2005correlation,supplemental}.

We add the separately computed $e$-ph and $e$-$e$ contributions to give the total scattering rate;
this is called Matthiessen's rule, which has broad experimental support.
There are few studies considering corrections beyond Matthiessen's rule.
Using the DFT+DMFT, $e$-ph coupling in FeSe was found to be enhanced due to electron correlations \cite{mandal2014strong},
and this has recently been verified experimentally \cite{gerber2017femtosecond}.
For pure hcp Fe, electron correlations may also change the strength of $e$-ph coupling but probably not strongly.
In recent work by Hausoel $et\ al.$ \cite{hausoel2017local},
the authors report on DMFT calculations of molecular dynamics snapshots of fcc Ni
and found that thermal disorder has only weak effects on electron correlations.
Matthiessen's rule is expected to be broken when in the saturation region when the resistivity approaches the Ioffe-Regel limit,
because there is essentially a minimum mean free path -- the nearest-neighbor distance \cite{gurvitch1981ioffe}.
Previous studies \cite{gunnarsson2003colloquium,rizzo2005transport} have shown that, for strongly-correlated systems,
resistivity can far exceed the Ioffe-Regel limit, corresponding to a very short mean free path.
Therefore, when $e$-ph and $e$-$e$ scattering contributions are comparable to each other, it may be suitable to
consider saturation effects only on the $e$-ph part and apply Matthiessen's rule after having considered saturation.
Since there is no evidence of the breakdown of Matthiessen's rule when considering $e$-$e$ and other scattering mechanisms,
we assume the applicability of Matthiessen's rule.

Using this approximation, we compute the total thermal conductivity $\kappa_{tot}\,=\,[\kappa_{e-ph}^{-1} + \kappa^{-1}_{e-e}]^{-1}$,
as in Ref. \cite{desjarlais2017density} for hydrogen plasma under extreme conditions.
The ionic part of thermal conductivity is neglected, since it is much smaller than the electronic part in metals.
As pointed out in Ref. \cite{pourovskii2017electron}, we observe relaxation time for the $e$-$e$ scattering $\tau_{e-e}$ being energy dependent,
although we disagree with their claim of iron being a simple Fermi liquid at high temperatures \cite{supplemental}.
Unlike $\rho_{e-e}$ being insensitive to the energy dependence of $\tau_{e-e}$, $\kappa_{e-e}$ can be considerably modified by its energy dependence.
We find that the Lorenz number for the $e$-$e$ scattering, $L_{e-e}$, is reduced from the conventional one $L_0$ by 20\%--45\%	, or
1.4--2.0$\times$10$^{-8}$ W$\Omega$K$^{-2}$, depending on the temperature and pressure.
This leads to a Lorenz number for $\rho_{tot}$ and $\kappa_{tot}$ of 2.10--2.15$\times$10$^{-8}$ W$\Omega$K$^{-2}$ at Earth's outer core conditions, from the core-mantle boundary (CMB, $P$ = 136 GPa and $T$ = 4000 K) to the inner core boundary (ICB, $P$ = 330 GPa and $T$ = 6000 K).

\begin{figure}[t]
\includegraphics[width=1.0\columnwidth]{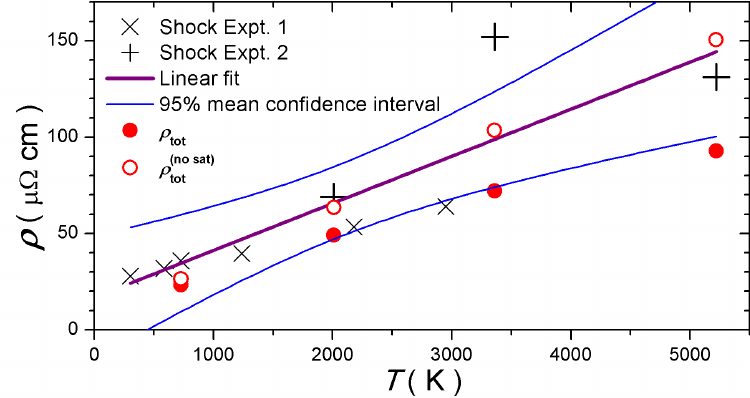}
\caption{\label{rho_hugoniot} Resistivity along the Hugoniot from shock data.
$\times$ are from Ref. \cite{keeler1969electrical} and + are from Ref. \cite{bi2002electrical}.
$\rho_{tot}=\rho_{e-ph}+\rho_{e-e}$, where $\rho_{e-ph}$ considers resistivity saturation effects using Eq. (\ref{eq:sat}),
and $\rho_{e-e}$ is electrical resistivity due to the electron-electron scattering.
$\rho^{(no\ sat)}_{tot}=\rho_B+\rho_{e-e}$, where $\rho_B$ is from the Boltzmann theory and does not consider resistivity saturation.
The purple line is the linear fit of the shock compression data.
The blue lines are the 95\% mean confidence interval.}
\end{figure}

\begin{figure}[t]
\includegraphics[width=1.0\columnwidth]{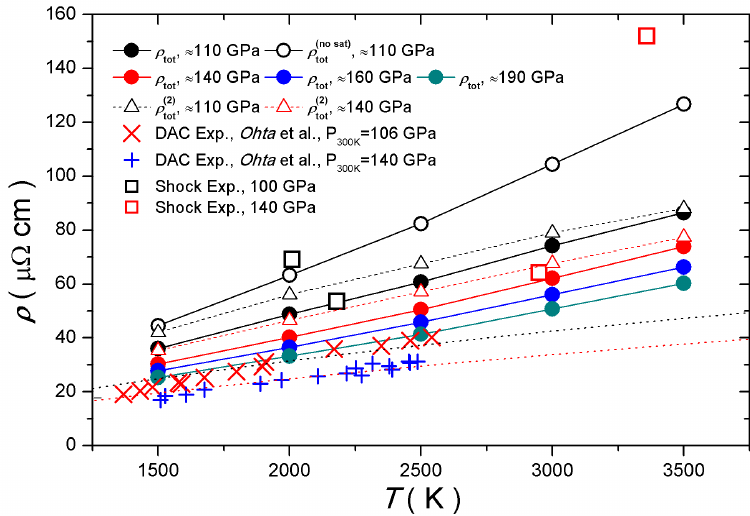}
\caption{\label{rho_106GPa} Calculated resistivity at fixed atomic volumes -- 57.9, 54.9, 53.3 and 51.6 bohr$^3$, corresponding to
pressures of about 110, 140, 160 and 190 GPa at 2500 K \cite{dewaele2006quasihydrostatic,sha2010first},
compared with shock data and experimental data by Ohta $et\ al.$ \cite{ohta2016experimental}.
$\rho_{tot}$ and $\rho_{tot}^{(no\ sat)}$ are total resistivity considering and not considering resistivity saturation effects, respectively.
$\rho^{(2)}_{tot}$ is the total resistivity to which the $e$-ph contribution is calculated using the KKR-CPA and the Kubo-Greenwood formula.
The dotted lines are their fits to the diamond anvil cell (DAC) data.}
\end{figure}

Our computed values of resistivity along the Hugoniot agree with the shock data from the experiments
\cite{keeler1969electrical,bi2002electrical} within the scatter (Fig. \ref{rho_hugoniot}).
We compared our computed isotropically averaged resistivity
at conditions close to the CMB ones with diamond anvil cell (DAC) data \cite{ohta2016experimental} (Fig. \ref{rho_106GPa}).
The computed resistivity is anisotropic, with $\rho_a\,/\,\rho_c=1.3$.
Using the BoltzTraP code \cite{madsen2006boltztrap}, we estimate $\rho_{sat}$ to be about 143 $\mu\Omega$cm at $V$ = 47.8 bohr$^3$/atom \cite{supplemental},
a bit lower than the estimate by Gomi $et\ al.$ \cite{gomi2013high}.
Our resistivities are somewhat higher than the experimental data, but broadly consistent,
considering the possibility of preferred orientation in the DAC experiments, temperature gradients,
and the large size of the probe wires compared with the sample.
Our calculations show a quasilinear relation between the total electrical resistivity and temperature for hcp Fe,
against the relation used in the fit of their experimental data by Ohta $et\ al.$,
where the slope of resistivity decreases with the temperature.
Including the $e$-$e$ contribution, the absolute value and the slope of the total resistivity become larger,
making the total resistivity more linear with the temperature.
Applying such a quasi-linear relation for extrapolation of experiments will increase their electrical resistivity at higher temperatures.

\begin{figure}[t]
\includegraphics[width=1.0\columnwidth]{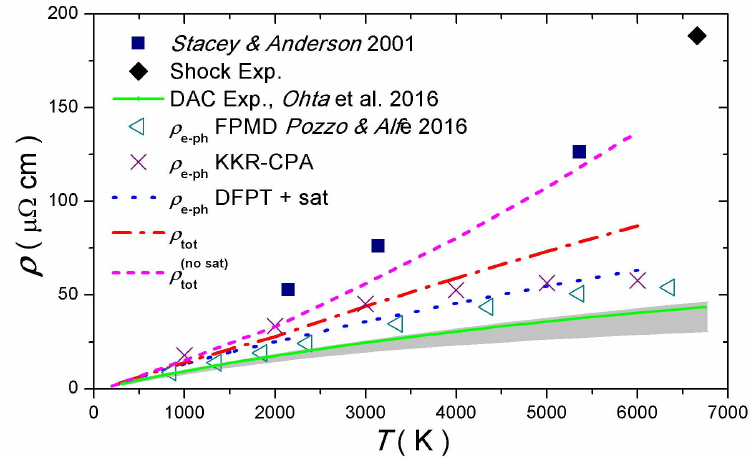}
\caption{\label{rho_v47.8} Resistivity versus temperature of hcp Fe at Earth's inner core density.
$\rho_{tot}$ and $\rho_{tot}^{(no\ sat)}$ are our calculated total resistivity considering and not considering resistivity saturation effects, respectively.
The crosses are results using the KKR-CPA which naturally includes saturation effects.
The open triangle is FPMD results \cite{pozzo2016saturation}, where there is no $e$-$e$ contribution, at a little higher density.
The navy squares are extrapolations to this density using the systematics of
Stacey and Anderson \cite{stacey2001electrical} based on the melting curve (which has no fundamental justification) .
The dark diamond is an interpolation to this density and an extrapolation to the 6658 K temperature
of previous shock compression results \cite{keeler1969electrical,bi2002electrical}.
The green line is from Ref. \cite{ohta2016experimental},
and is the extrapolation of their experimental data.}
\end{figure}

We compared our calculated resistivity
with previous theoretical and experimental results at the inner core density of iron
(13.04 g cm$^{-3}$, atomic volume of 47.8 atomic units  = 7.083 \AA$^3$) (Fig. \ref{rho_v47.8}).
Our $e$-ph results are slightly higher but in general agreement with the FPMD results \cite{pozzo2016saturation}.
Our total resistivity is in quite poor agreement with the extrapolation values of DAC data \cite{ohta2016experimental},
consistent with their extrapolation not being accurate from overestimating saturation effects.
In addition to possible experimental errors of temperatures in their measurements, the disagreement may be also due to
their use of smaller $\rho_{sat}$ and the neglect of the $e$-$e$ scattering in the temperature dependence
of the resistivity in their extrapolation.
At higher temperatures, the $e$-$e$ scattering becomes more important and reaches about 35\% of the $e$-ph value at Earth's core conditions.

\begin{figure}[t]
\includegraphics[width=1.0\columnwidth]{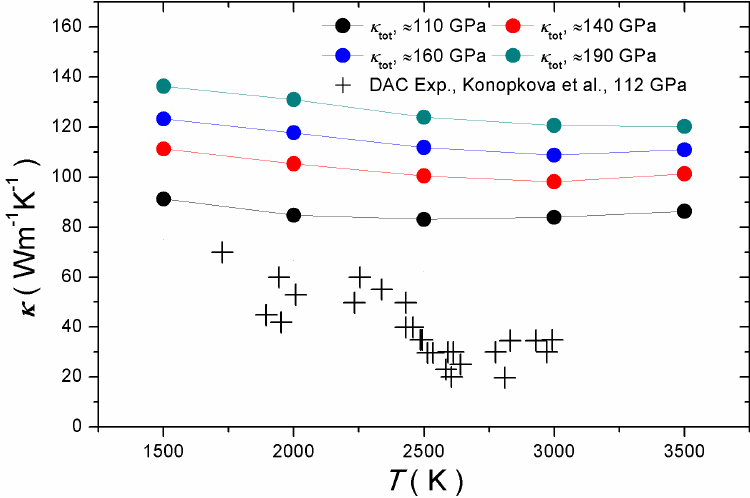}
\caption{\label{kappa_112GPa} Our calculated thermal conductivity of hcp Fe
at fixed atomic volumes -- 57.9, 54.9, 53.3 and 51.6 bohr$^3$, corresponding to
pressures about 110, 140, 160 and 190 GPa at 2500 K \cite{dewaele2006quasihydrostatic,sha2010first},
compared with experimental data \cite{konopkova2016direct}.
$\kappa_{tot}$ is the total thermal conductivity and is equal to
$\kappa_{tot} = [\kappa_{e-ph}^{-1} + \kappa^{-1}_{e-e}]^{-1}$.}
\end{figure}

\begin{figure}[t]
\includegraphics[width=1.0\columnwidth]{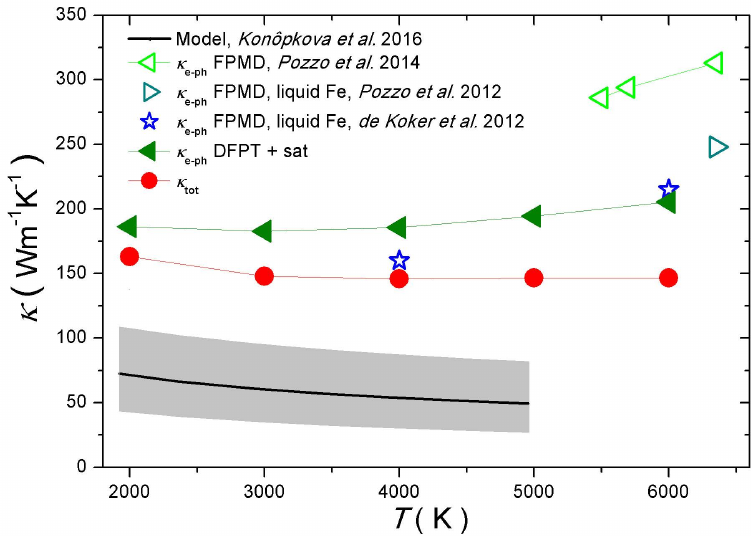}
\caption{\label{kappa_v47.8} Calculated thermal conductivity of solid hcp Fe at Earth's core density,
13.04 g cm$^{-3}$, or atomic volume 47.8 bohr$^3$. At 6000 K, the corresponding pressure is about 305 GPa.
$\kappa_{tot}$ is our calculated total thermal conductivity using the DFPT and DMFT.
The black line is the extrapolation model at 330 GPa
based on the experimental data at 112 GPa \cite{konopkova2016direct}.
A uncertainty envelope of the model (gray) is given according to
our guess that their experimental data at 2000 K and the extrapolation to
pressure 330 GPa may both have errors of 10\%--30\% and the thermal conductivity may decrease slower than $T^{-0.5}$, as assumed in their model.
The green open triangles are the theoretical values of thermal conductivity at 329 GPa due to only $e$-ph scattering
using FPMD \cite{pozzo2014thermal}, which includes saturation effects consistent with system size and $k$-point sampling.
The dark cyan open triangle and blue open stars are the theoretical thermal conductivity of liquid Fe due to only $e$-ph scattering using FPMD \cite{pozzo2012thermal,de2012electrical}.}
\end{figure}

We compared our calculated thermal conductivity of hcp Fe at conditions close to CMB ones
with experimental data \cite{konopkova2016direct} and find that the agreement is good at 2000 K, but becomes poor above 2400 K (Fig. \ref{kappa_112GPa}).

We calculate thermal conductivity at inner core density
and obtain the theoretical electronic part of the thermal conductivity of pure solid iron, about 147 W m$^{-1}$ K$^{-1}$, at inner core conditions (Fig. \ref{rho_v47.8}).
The agreement with the extrapolation model at 330 GPa based on the experimental data at 112 GPa \cite{konopkova2016direct} is quite poor.
Except for possible errors in the measurements, and various assumptions made in our calculations,
a possible reason for the disagreement may be that their extrapolation method is not accurate,
since only the variation of thermal conductivity as a function of pressure and temperature due to $e$-ph scattering is considered.

Using previous estimates of thermal conductivity, i.e. with the theoretical thermal conductivity of
liquid Fe-Si or Fe-O alloy (pure liquid Fe) at CMB conditions, of about 100 (140) W m$^{-1}$ K$^{-1}$,
the heat loss from the core to the mantle by conduction is estimated to be 15 TW \cite{pozzo2012thermal}.
The total heat from the core is estimated to 8-16 TW \cite{lay2008core,wu2011statistical}
so that the conventional thermal convection geodynamo model would probably fail.
At CMB conditions, we find $\kappa\approx97$ W m$^{-1}$ K$^{-1}$ for pure solid hcp Fe (Fig. \ref{kappa_112GPa}).
Earth's outer core contains light elements of the order of 20\%, and light elements will probably decrease the $e$-ph scattering contribution to thermal conductivity
by 10\%--30\% \cite{pozzo2012thermal,de2012electrical,gomi2013high}.
In addition, melting will decrease the density and may further lead to an $\approx$10\% reduction
of both the $e$-ph contribution \cite{pozzo2014thermal} and the $e$-$e$ contribution.
The thermal conductivity would accordingly be about 77 W m$^{-1}$ K$^{-1}$.
The corresponding heat conduction down the core adiabat will be about 9--12 TW,
depending on the choice of core parameters, e.g., specific heat capacity, CMB temperature, etc.
\cite{stacey2007revised,pozzo2012thermal,nimmo2015treatise}.\par

Another candidate phase of solid Fe alloy at Earth's core conditions is bcc \cite{belonoshko2017stabilization,Vocadlo2008},
which is dynamically unstable, so we cannot fully apply DFPT to compute transport properties; for completeness
we estimate the $e$-ph scattering contribution of bcc Fe neglecting the unstable modes
and calculate the $e$-$e$ scattering contribution using DFT+DMFT.
We find that the total resistivity of bcc Fe is different from that of hcp Fe by several percent.
All of the above results are for solid iron, but experiments on many materials show that melting typically increases resistivity by 5\%--10\%.
We computed the effects of melting on iron by applying DFT+DMFT to snapshots of liquid Fe from first-principles molecular dynamics \cite{supplemental}
and find that scattering rates due to $e$-$e$ scattering around the chemical potential are about 20\%--40\% larger than those of solid Fe at similar conditions,
and that the thermal conductivity of liquid iron at Earth's core conditions is reduced by about 10\% from solid iron.
A full discussion of results for liquid iron and its alloys will be discussed in another work, as it is a separate ongoing study.

Our final estimates for the thermal conductivity for pure solid hcp iron at Earth's inner core conditions is $\kappa=$ 147 W m$^{-1}$ K$^{-1}$, 97 W m$^{-1}$ K$^{-1}$ at the core-mantle boundary temperature and pressure, and 77$\pm$10 for liquid iron alloy in the outer core.  This is consistent with a thermally convection driven dynamo throughout Earth history, which requires $\kappa=$100 W m$^{-1}$ K$^{-1}$ for  a CMB heat flow of about 15 TW and temperature  $T_{CMB}$ = 4000 K \cite{Driscoll2014}.

We thank Peter Driscoll, Jung-Fu Lin, and Youjun Zhang for helpful discussions.
This work was supported by the European Research Council Advanced Grant ToMCaT and by the Carnegie Institution for Science.
The authors gratefully acknowledge the Gauss Centre for Supercomputing (GCS) e.V. for funding this project
by providing computing time on the GCS Supercomputer SuperMUC at Leibniz Supercomputing Centre (LRZ).
P. Z. acknowledges support of National Science Foundation of China (Grant No. 11604255).
J. M. further thanks for the support from the Computational and Experimental Design of Advanced Materials with New Functionalities
(CEDAMNF) project (CZ.02.1.01/0.0/0.0/15\_003/0000358) of Czech ministerium MSMT.

\nocite{*}

\bibliography{Manuscript_hcpFe}

\providecommand{\noopsort}[1]{}\providecommand{\singleletter}[1]{#1}%
\begin{thebibliography}{65}%
\makeatletter
\providecommand \@ifxundefined [1]{%
 \@ifx{#1\undefined}
}%
\providecommand \@ifnum [1]{%
 \ifnum #1\expandafter \@firstoftwo
 \else \expandafter \@secondoftwo
 \fi
}%
\providecommand \@ifx [1]{%
 \ifx #1\expandafter \@firstoftwo
 \else \expandafter \@secondoftwo
 \fi
}%
\providecommand \natexlab [1]{#1}%
\providecommand \enquote  [1]{``#1''}%
\providecommand \bibnamefont  [1]{#1}%
\providecommand \bibfnamefont [1]{#1}%
\providecommand \citenamefont [1]{#1}%
\providecommand \href@noop [0]{\@secondoftwo}%
\providecommand \href [0]{\begingroup \@sanitize@url \@href}%
\providecommand \@href[1]{\@@startlink{#1}\@@href}%
\providecommand \@@href[1]{\endgroup#1\@@endlink}%
\providecommand \@sanitize@url [0]{\catcode `\\12\catcode `\$12\catcode
  `\&12\catcode `\#12\catcode `\^12\catcode `\_12\catcode `\%12\relax}%
\providecommand \@@startlink[1]{}%
\providecommand \@@endlink[0]{}%
\providecommand \url  [0]{\begingroup\@sanitize@url \@url }%
\providecommand \@url [1]{\endgroup\@href {#1}{\urlprefix }}%
\providecommand \urlprefix  [0]{URL }%
\providecommand \Eprint [0]{\href }%
\providecommand \doibase [0]{http://dx.doi.org/}%
\providecommand \selectlanguage [0]{\@gobble}%
\providecommand \bibinfo  [0]{\@secondoftwo}%
\providecommand \bibfield  [0]{\@secondoftwo}%
\providecommand \translation [1]{[#1]}%
\providecommand \BibitemOpen [0]{}%
\providecommand \bibitemStop [0]{}%
\providecommand \bibitemNoStop [0]{.\EOS\space}%
\providecommand \EOS [0]{\spacefactor3000\relax}%
\providecommand \BibitemShut  [1]{\csname bibitem#1\endcsname}%
\let\auto@bib@innerbib\@empty
\bibitem [{\citenamefont {Labrosse}(2003)}]{labrosse2003thermal}%
  \BibitemOpen
  \bibfield  {author} {\bibinfo {author} {\bibfnamefont {S.}~\bibnamefont
  {Labrosse}},\ }\href@noop {} {\bibfield  {journal} {\bibinfo  {journal}
  {Physics of the Earth and Planetary Interiors}\ }\textbf {\bibinfo {volume}
  {140}},\ \bibinfo {pages} {127} (\bibinfo {year} {2003})}\BibitemShut
  {NoStop}%
\bibitem [{\citenamefont {Nimmo}(2015)}]{nimmo2015treatise}%
  \BibitemOpen
  \bibfield  {author} {\bibinfo {author} {\bibfnamefont {F.}~\bibnamefont
  {Nimmo}},\ }\href@noop {} {\emph {\bibinfo {title} {Treatise on geophysics:
  Energetics of the Core}}}\ (\bibinfo  {publisher} {Elsevier},\ \bibinfo
  {year} {2015})\BibitemShut {NoStop}%
\bibitem [{\citenamefont {Stacey}\ and\ \citenamefont
  {Loper}(2007)}]{stacey2007revised}%
  \BibitemOpen
  \bibfield  {author} {\bibinfo {author} {\bibfnamefont {F.}~\bibnamefont
  {Stacey}}\ and\ \bibinfo {author} {\bibfnamefont {D.}~\bibnamefont {Loper}},\
  }\href@noop {} {\bibfield  {journal} {\bibinfo  {journal} {Physics of the
  Earth and Planetary Interiors}\ }\textbf {\bibinfo {volume} {161}},\ \bibinfo
  {pages} {13} (\bibinfo {year} {2007})}\BibitemShut {NoStop}%
\bibitem [{\citenamefont {Sha}\ and\ \citenamefont
  {Cohen}(2011)}]{sha2011first}%
  \BibitemOpen
  \bibfield  {author} {\bibinfo {author} {\bibfnamefont {X.}~\bibnamefont
  {Sha}}\ and\ \bibinfo {author} {\bibfnamefont {R.~E.}\ \bibnamefont
  {Cohen}},\ }\href@noop {} {\bibfield  {journal} {\bibinfo  {journal} {Journal
  of Physics: Condensed Matter}\ }\textbf {\bibinfo {volume} {23}},\ \bibinfo
  {pages} {075401} (\bibinfo {year} {2011})}\BibitemShut {NoStop}%
\bibitem [{\citenamefont {de~Koker}\ \emph {et~al.}(2012)\citenamefont
  {de~Koker}, \citenamefont {Steinle-Neumann},\ and\ \citenamefont
  {Vl{\v{c}}ek}}]{de2012electrical}%
  \BibitemOpen
  \bibfield  {author} {\bibinfo {author} {\bibfnamefont {N.}~\bibnamefont
  {de~Koker}}, \bibinfo {author} {\bibfnamefont {G.}~\bibnamefont
  {Steinle-Neumann}}, \ and\ \bibinfo {author} {\bibfnamefont {V.}~\bibnamefont
  {Vl{\v{c}}ek}},\ }\href@noop {} {\bibfield  {journal} {\bibinfo  {journal}
  {Proceedings of the National Academy of Sciences}\ }\textbf {\bibinfo
  {volume} {109}},\ \bibinfo {pages} {4070} (\bibinfo {year}
  {2012})}\BibitemShut {NoStop}%
\bibitem [{\citenamefont {Pozzo}\ \emph {et~al.}(2012)\citenamefont {Pozzo},
  \citenamefont {Davies}, \citenamefont {Gubbins},\ and\ \citenamefont
  {Alfe}}]{pozzo2012thermal}%
  \BibitemOpen
  \bibfield  {author} {\bibinfo {author} {\bibfnamefont {M.}~\bibnamefont
  {Pozzo}}, \bibinfo {author} {\bibfnamefont {C.}~\bibnamefont {Davies}},
  \bibinfo {author} {\bibfnamefont {D.}~\bibnamefont {Gubbins}}, \ and\
  \bibinfo {author} {\bibfnamefont {D.}~\bibnamefont {Alfe}},\ }\href@noop {}
  {\bibfield  {journal} {\bibinfo  {journal} {Nature}\ }\textbf {\bibinfo
  {volume} {485}},\ \bibinfo {pages} {355} (\bibinfo {year}
  {2012})}\BibitemShut {NoStop}%
\bibitem [{\citenamefont {Zhang}\ \emph {et~al.}(2015)\citenamefont {Zhang},
  \citenamefont {Cohen},\ and\ \citenamefont {Haule}}]{zhang2015effects}%
  \BibitemOpen
  \bibfield  {author} {\bibinfo {author} {\bibfnamefont {P.}~\bibnamefont
  {Zhang}}, \bibinfo {author} {\bibfnamefont {R.}~\bibnamefont {Cohen}}, \ and\
  \bibinfo {author} {\bibfnamefont {K.}~\bibnamefont {Haule}},\ }\href@noop {}
  {\bibfield  {journal} {\bibinfo  {journal} {Nature}\ }\textbf {\bibinfo
  {volume} {517}},\ \bibinfo {pages} {605} (\bibinfo {year}
  {2015})}\BibitemShut {NoStop}%
\bibitem [{\citenamefont {Secco}(2017)}]{secco2017thermal}%
  \BibitemOpen
  \bibfield  {author} {\bibinfo {author} {\bibfnamefont {R.~A.}\ \bibnamefont
  {Secco}},\ }\href@noop {} {\bibfield  {journal} {\bibinfo  {journal} {Physics
  of the Earth and Planetary Interiors}\ }\textbf {\bibinfo {volume} {265}},\
  \bibinfo {pages} {23} (\bibinfo {year} {2017})}\BibitemShut {NoStop}%
\bibitem [{\citenamefont {Keeler}\ and\ \citenamefont
  {Mitchell}(1969)}]{keeler1969electrical}%
  \BibitemOpen
  \bibfield  {author} {\bibinfo {author} {\bibfnamefont {R.}~\bibnamefont
  {Keeler}}\ and\ \bibinfo {author} {\bibfnamefont {A.}~\bibnamefont
  {Mitchell}},\ }\href@noop {} {\bibfield  {journal} {\bibinfo  {journal}
  {Solid State Communications}\ }\textbf {\bibinfo {volume} {7}},\ \bibinfo
  {pages} {271} (\bibinfo {year} {1969})}\BibitemShut {NoStop}%
\bibitem [{\citenamefont {Gomi}\ \emph {et~al.}(2013)\citenamefont {Gomi},
  \citenamefont {Ohta}, \citenamefont {Hirose}, \citenamefont {Labrosse},
  \citenamefont {Caracas}, \citenamefont {Verstraete},\ and\ \citenamefont
  {Hernlund}}]{gomi2013high}%
  \BibitemOpen
  \bibfield  {author} {\bibinfo {author} {\bibfnamefont {H.}~\bibnamefont
  {Gomi}}, \bibinfo {author} {\bibfnamefont {K.}~\bibnamefont {Ohta}}, \bibinfo
  {author} {\bibfnamefont {K.}~\bibnamefont {Hirose}}, \bibinfo {author}
  {\bibfnamefont {S.}~\bibnamefont {Labrosse}}, \bibinfo {author}
  {\bibfnamefont {R.}~\bibnamefont {Caracas}}, \bibinfo {author} {\bibfnamefont
  {M.~J.}\ \bibnamefont {Verstraete}}, \ and\ \bibinfo {author} {\bibfnamefont
  {J.~W.}\ \bibnamefont {Hernlund}},\ }\href@noop {} {\bibfield  {journal}
  {\bibinfo  {journal} {Physics of the Earth and Planetary Interiors}\ }\textbf
  {\bibinfo {volume} {224}},\ \bibinfo {pages} {88} (\bibinfo {year}
  {2013})}\BibitemShut {NoStop}%
\bibitem [{\citenamefont {Ohta}\ \emph {et~al.}(2016)\citenamefont {Ohta},
  \citenamefont {Kuwayama}, \citenamefont {Hirose}, \citenamefont {Shimizu},\
  and\ \citenamefont {Ohishi}}]{ohta2016experimental}%
  \BibitemOpen
  \bibfield  {author} {\bibinfo {author} {\bibfnamefont {K.}~\bibnamefont
  {Ohta}}, \bibinfo {author} {\bibfnamefont {Y.}~\bibnamefont {Kuwayama}},
  \bibinfo {author} {\bibfnamefont {K.}~\bibnamefont {Hirose}}, \bibinfo
  {author} {\bibfnamefont {K.}~\bibnamefont {Shimizu}}, \ and\ \bibinfo
  {author} {\bibfnamefont {Y.}~\bibnamefont {Ohishi}},\ }\href@noop {}
  {\bibfield  {journal} {\bibinfo  {journal} {Nature}\ }\textbf {\bibinfo
  {volume} {534}},\ \bibinfo {pages} {95} (\bibinfo {year} {2016})}\BibitemShut
  {NoStop}%
\bibitem [{\citenamefont {Jonson}\ and\ \citenamefont
  {Mahan}(1980)}]{jonson1980mott}%
  \BibitemOpen
  \bibfield  {author} {\bibinfo {author} {\bibfnamefont {M.}~\bibnamefont
  {Jonson}}\ and\ \bibinfo {author} {\bibfnamefont {G.~D.}\ \bibnamefont
  {Mahan}},\ }\href@noop {} {\bibfield  {journal} {\bibinfo  {journal}
  {Physical Review B}\ }\textbf {\bibinfo {volume} {21}},\ \bibinfo {pages}
  {4223} (\bibinfo {year} {1980})}\BibitemShut {NoStop}%
\bibitem [{\citenamefont {Kon{\^o}pkov{\'a}}\ \emph {et~al.}(2016)\citenamefont
  {Kon{\^o}pkov{\'a}}, \citenamefont {McWilliams}, \citenamefont
  {G{\'o}mez-P{\'e}rez},\ and\ \citenamefont
  {Goncharov}}]{konopkova2016direct}%
  \BibitemOpen
  \bibfield  {author} {\bibinfo {author} {\bibfnamefont {Z.}~\bibnamefont
  {Kon{\^o}pkov{\'a}}}, \bibinfo {author} {\bibfnamefont {R.~S.}\ \bibnamefont
  {McWilliams}}, \bibinfo {author} {\bibfnamefont {N.}~\bibnamefont
  {G{\'o}mez-P{\'e}rez}}, \ and\ \bibinfo {author} {\bibfnamefont {A.~F.}\
  \bibnamefont {Goncharov}},\ }\href@noop {} {\bibfield  {journal} {\bibinfo
  {journal} {Nature}\ }\textbf {\bibinfo {volume} {534}},\ \bibinfo {pages}
  {99} (\bibinfo {year} {2016})}\BibitemShut {NoStop}%
\bibitem [{\citenamefont {O'rourke}\ and\ \citenamefont
  {Stevenson}(2016)}]{o2016powering}%
  \BibitemOpen
  \bibfield  {author} {\bibinfo {author} {\bibfnamefont {J.~G.}\ \bibnamefont
  {O'rourke}}\ and\ \bibinfo {author} {\bibfnamefont {D.~J.}\ \bibnamefont
  {Stevenson}},\ }\href@noop {} {\bibfield  {journal} {\bibinfo  {journal}
  {Nature}\ }\textbf {\bibinfo {volume} {529}},\ \bibinfo {pages} {387}
  (\bibinfo {year} {2016})}\BibitemShut {NoStop}%
\bibitem [{\citenamefont {Badro}\ \emph {et~al.}(2016)\citenamefont {Badro},
  \citenamefont {Siebert},\ and\ \citenamefont {Nimmo}}]{badro2016early}%
  \BibitemOpen
  \bibfield  {author} {\bibinfo {author} {\bibfnamefont {J.}~\bibnamefont
  {Badro}}, \bibinfo {author} {\bibfnamefont {J.}~\bibnamefont {Siebert}}, \
  and\ \bibinfo {author} {\bibfnamefont {F.}~\bibnamefont {Nimmo}},\
  }\href@noop {} {\bibfield  {journal} {\bibinfo  {journal} {Nature}\ }\textbf
  {\bibinfo {volume} {536}},\ \bibinfo {pages} {326} (\bibinfo {year}
  {2016})}\BibitemShut {NoStop}%
\bibitem [{\citenamefont {Allen}(1978)}]{allen1978new}%
  \BibitemOpen
  \bibfield  {author} {\bibinfo {author} {\bibfnamefont {P.}~\bibnamefont
  {Allen}},\ }\href@noop {} {\bibfield  {journal} {\bibinfo  {journal}
  {Physical Review B}\ }\textbf {\bibinfo {volume} {17}},\ \bibinfo {pages}
  {3725} (\bibinfo {year} {1978})}\BibitemShut {NoStop}%
\bibitem [{\citenamefont {Gonze}\ \emph {et~al.}(2016)\citenamefont {Gonze},
  \citenamefont {Jollet}, \citenamefont {Araujo} \emph {et~al.}}]{gonze2016}%
  \BibitemOpen
  \bibfield  {author} {\bibinfo {author} {\bibfnamefont {X.}~\bibnamefont
  {Gonze}}, \bibinfo {author} {\bibfnamefont {F.}~\bibnamefont {Jollet}},
  \bibinfo {author} {\bibfnamefont {F.~A.}\ \bibnamefont {Araujo}},  \emph
  {et~al.},\ }\href@noop {} {\bibfield  {journal} {\bibinfo  {journal}
  {Computer Physics Communications}\ }\textbf {\bibinfo {volume} {205}},\
  \bibinfo {pages} {106} (\bibinfo {year} {2016})}\BibitemShut {NoStop}%
\bibitem [{sup()}]{supplemental}%
  \BibitemOpen
  \href@noop {} {\bibinfo  {journal} {See Supplemental Material for more
  technical details and theoretical results, which additionally includes Refs.
  [54-65]}\ }\BibitemShut {NoStop}%
\bibitem [{\citenamefont {Wiesmann}\ \emph {et~al.}(1977)\citenamefont
  {Wiesmann}, \citenamefont {Gurvitch}, \citenamefont {Lutz}, \citenamefont
  {Ghosh}, \citenamefont {Schwarz}, \citenamefont {Strongin}, \citenamefont
  {Allen},\ and\ \citenamefont {Halley}}]{wiesmann1977simple}%
  \BibitemOpen
\bibfield  {journal} {  }\bibfield  {author} {\bibinfo {author} {\bibfnamefont
  {H.}~\bibnamefont {Wiesmann}}, \bibinfo {author} {\bibfnamefont
  {M.}~\bibnamefont {Gurvitch}}, \bibinfo {author} {\bibfnamefont
  {H.}~\bibnamefont {Lutz}}, \bibinfo {author} {\bibfnamefont {A.}~\bibnamefont
  {Ghosh}}, \bibinfo {author} {\bibfnamefont {B.}~\bibnamefont {Schwarz}},
  \bibinfo {author} {\bibfnamefont {M.}~\bibnamefont {Strongin}}, \bibinfo
  {author} {\bibfnamefont {P.}~\bibnamefont {Allen}}, \ and\ \bibinfo {author}
  {\bibfnamefont {J.}~\bibnamefont {Halley}},\ }\href@noop {} {\bibfield
  {journal} {\bibinfo  {journal} {Physical Review Letters}\ }\textbf {\bibinfo
  {volume} {38}},\ \bibinfo {pages} {782} (\bibinfo {year} {1977})}\BibitemShut
  {NoStop}%
\bibitem [{\citenamefont {Allen}\ and\ \citenamefont
  {Chakraborty}(1981)}]{allen1981infrared}%
  \BibitemOpen
  \bibfield  {author} {\bibinfo {author} {\bibfnamefont {P.}~\bibnamefont
  {Allen}}\ and\ \bibinfo {author} {\bibfnamefont {B.}~\bibnamefont
  {Chakraborty}},\ }\href@noop {} {\bibfield  {journal} {\bibinfo  {journal}
  {Physical Review B}\ }\textbf {\bibinfo {volume} {23}},\ \bibinfo {pages}
  {4815} (\bibinfo {year} {1981})}\BibitemShut {NoStop}%
\bibitem [{\citenamefont {Schiller}\ and\ \citenamefont
  {Langowski}(1982)}]{schiller1982influence}%
  \BibitemOpen
  \bibfield  {author} {\bibinfo {author} {\bibfnamefont {W.}~\bibnamefont
  {Schiller}}\ and\ \bibinfo {author} {\bibfnamefont {G.}~\bibnamefont
  {Langowski}},\ }\href@noop {} {\bibfield  {journal} {\bibinfo  {journal}
  {Journal of Physics F: Metal Physics}\ }\textbf {\bibinfo {volume} {12}},\
  \bibinfo {pages} {449} (\bibinfo {year} {1982})}\BibitemShut {NoStop}%
\bibitem [{\citenamefont {Werman}\ \emph {et~al.}(2017)\citenamefont {Werman},
  \citenamefont {Kivelson},\ and\ \citenamefont {Berg}}]{werman2017non}%
  \BibitemOpen
  \bibfield  {author} {\bibinfo {author} {\bibfnamefont {Y.}~\bibnamefont
  {Werman}}, \bibinfo {author} {\bibfnamefont {S.~A.}\ \bibnamefont
  {Kivelson}}, \ and\ \bibinfo {author} {\bibfnamefont {E.}~\bibnamefont
  {Berg}},\ }\href@noop {} {\bibfield  {journal} {\bibinfo  {journal} {npj
  Quantum Materials}\ }\textbf {\bibinfo {volume} {2}},\ \bibinfo {pages} {7}
  (\bibinfo {year} {2017})}\BibitemShut {NoStop}%
\bibitem [{\citenamefont {Ebert}\ \emph {et~al.}(2015)\citenamefont {Ebert},
  \citenamefont {Mankovsky}, \citenamefont {Chadova}, \citenamefont {Polesya},
  \citenamefont {Minar},\ and\ \citenamefont
  {Koedderitzsch}}]{ebert2015calculating}%
  \BibitemOpen
  \bibfield  {author} {\bibinfo {author} {\bibfnamefont {H.}~\bibnamefont
  {Ebert}}, \bibinfo {author} {\bibfnamefont {S.}~\bibnamefont {Mankovsky}},
  \bibinfo {author} {\bibfnamefont {K.}~\bibnamefont {Chadova}}, \bibinfo
  {author} {\bibfnamefont {S.}~\bibnamefont {Polesya}}, \bibinfo {author}
  {\bibfnamefont {J.}~\bibnamefont {Minar}}, \ and\ \bibinfo {author}
  {\bibfnamefont {D.}~\bibnamefont {Koedderitzsch}},\ }\href@noop {} {\bibfield
   {journal} {\bibinfo  {journal} {Physical Review B}\ }\textbf {\bibinfo
  {volume} {91}},\ \bibinfo {pages} {165132} (\bibinfo {year}
  {2015})}\BibitemShut {NoStop}%
\bibitem [{\citenamefont {Ebert}\ \emph {et~al.}(2011)\citenamefont {Ebert},
  \citenamefont {Koedderitzsch},\ and\ \citenamefont
  {Minar}}]{ebert2011calculating}%
  \BibitemOpen
  \bibfield  {author} {\bibinfo {author} {\bibfnamefont {H.}~\bibnamefont
  {Ebert}}, \bibinfo {author} {\bibfnamefont {D.}~\bibnamefont
  {Koedderitzsch}}, \ and\ \bibinfo {author} {\bibfnamefont {J.}~\bibnamefont
  {Minar}},\ }\href@noop {} {\bibfield  {journal} {\bibinfo  {journal} {Reports
  on Progress in Physics}\ }\textbf {\bibinfo {volume} {74}},\ \bibinfo {pages}
  {096501} (\bibinfo {year} {2011})}\BibitemShut {NoStop}%
\bibitem [{\citenamefont {Dewaele}\ \emph {et~al.}(2006)\citenamefont
  {Dewaele}, \citenamefont {Loubeyre}, \citenamefont {Occelli}, \citenamefont
  {Mezouar}, \citenamefont {Dorogokupets},\ and\ \citenamefont
  {Torrent}}]{dewaele2006quasihydrostatic}%
  \BibitemOpen
  \bibfield  {author} {\bibinfo {author} {\bibfnamefont {A.}~\bibnamefont
  {Dewaele}}, \bibinfo {author} {\bibfnamefont {P.}~\bibnamefont {Loubeyre}},
  \bibinfo {author} {\bibfnamefont {F.}~\bibnamefont {Occelli}}, \bibinfo
  {author} {\bibfnamefont {M.}~\bibnamefont {Mezouar}}, \bibinfo {author}
  {\bibfnamefont {P.~I.}\ \bibnamefont {Dorogokupets}}, \ and\ \bibinfo
  {author} {\bibfnamefont {M.}~\bibnamefont {Torrent}},\ }\href@noop {}
  {\bibfield  {journal} {\bibinfo  {journal} {Physical Review Letters}\
  }\textbf {\bibinfo {volume} {97}},\ \bibinfo {pages} {215504} (\bibinfo
  {year} {2006})}\BibitemShut {NoStop}%
\bibitem [{\citenamefont {Sha}\ and\ \citenamefont
  {Cohen}(2010)}]{sha2010first}%
  \BibitemOpen
  \bibfield  {author} {\bibinfo {author} {\bibfnamefont {X.}~\bibnamefont
  {Sha}}\ and\ \bibinfo {author} {\bibfnamefont {R.~E.}\ \bibnamefont
  {Cohen}},\ }\href@noop {} {\bibfield  {journal} {\bibinfo  {journal}
  {Physical Review B}\ }\textbf {\bibinfo {volume} {81}},\ \bibinfo {pages}
  {094105} (\bibinfo {year} {2010})}\BibitemShut {NoStop}%
\bibitem [{\citenamefont {Georges}\ \emph {et~al.}(1996)\citenamefont
  {Georges}, \citenamefont {Kotliar}, \citenamefont {Krauth},\ and\
  \citenamefont {Rozenberg}}]{georges1996dynamical}%
  \BibitemOpen
  \bibfield  {author} {\bibinfo {author} {\bibfnamefont {A.}~\bibnamefont
  {Georges}}, \bibinfo {author} {\bibfnamefont {G.}~\bibnamefont {Kotliar}},
  \bibinfo {author} {\bibfnamefont {W.}~\bibnamefont {Krauth}}, \ and\ \bibinfo
  {author} {\bibfnamefont {M.~J.}\ \bibnamefont {Rozenberg}},\ }\href@noop {}
  {\bibfield  {journal} {\bibinfo  {journal} {Reviews of Modern Physics}\
  }\textbf {\bibinfo {volume} {68}},\ \bibinfo {pages} {13} (\bibinfo {year}
  {1996})}\BibitemShut {NoStop}%
\bibitem [{\citenamefont {Kotliar}\ \emph {et~al.}(2006)\citenamefont
  {Kotliar}, \citenamefont {Savrasov}, \citenamefont {Haule}, \citenamefont
  {Oudovenko}, \citenamefont {Parcollet},\ and\ \citenamefont
  {Marianetti}}]{kotliar2006electronic}%
  \BibitemOpen
  \bibfield  {author} {\bibinfo {author} {\bibfnamefont {G.}~\bibnamefont
  {Kotliar}}, \bibinfo {author} {\bibfnamefont {S.~Y.}\ \bibnamefont
  {Savrasov}}, \bibinfo {author} {\bibfnamefont {K.}~\bibnamefont {Haule}},
  \bibinfo {author} {\bibfnamefont {V.~S.}\ \bibnamefont {Oudovenko}}, \bibinfo
  {author} {\bibfnamefont {O.}~\bibnamefont {Parcollet}}, \ and\ \bibinfo
  {author} {\bibfnamefont {C.}~\bibnamefont {Marianetti}},\ }\href@noop {}
  {\bibfield  {journal} {\bibinfo  {journal} {Reviews of Modern Physics}\
  }\textbf {\bibinfo {volume} {78}},\ \bibinfo {pages} {865} (\bibinfo {year}
  {2006})}\BibitemShut {NoStop}%
\bibitem [{\citenamefont {Haule}(2007)}]{haule2007quantum}%
  \BibitemOpen
  \bibfield  {author} {\bibinfo {author} {\bibfnamefont {K.}~\bibnamefont
  {Haule}},\ }\href@noop {} {\bibfield  {journal} {\bibinfo  {journal}
  {Physical Review B}\ }\textbf {\bibinfo {volume} {75}},\ \bibinfo {pages}
  {155113} (\bibinfo {year} {2007})}\BibitemShut {NoStop}%
\bibitem [{\citenamefont {Gull}\ \emph {et~al.}(2011)\citenamefont {Gull},
  \citenamefont {Millis}, \citenamefont {Lichtenstein}, \citenamefont
  {Rubtsov}, \citenamefont {Troyer},\ and\ \citenamefont
  {Werner}}]{gull2011continuous}%
  \BibitemOpen
  \bibfield  {author} {\bibinfo {author} {\bibfnamefont {E.}~\bibnamefont
  {Gull}}, \bibinfo {author} {\bibfnamefont {A.~J.}\ \bibnamefont {Millis}},
  \bibinfo {author} {\bibfnamefont {A.~I.}\ \bibnamefont {Lichtenstein}},
  \bibinfo {author} {\bibfnamefont {A.~N.}\ \bibnamefont {Rubtsov}}, \bibinfo
  {author} {\bibfnamefont {M.}~\bibnamefont {Troyer}}, \ and\ \bibinfo {author}
  {\bibfnamefont {P.}~\bibnamefont {Werner}},\ }\href@noop {} {\bibfield
  {journal} {\bibinfo  {journal} {Reviews of Modern Physics}\ }\textbf
  {\bibinfo {volume} {83}},\ \bibinfo {pages} {349} (\bibinfo {year}
  {2011})}\BibitemShut {NoStop}%
\bibitem [{\citenamefont {Haule}\ \emph {et~al.}(2010)\citenamefont {Haule},
  \citenamefont {Yee},\ and\ \citenamefont {Kim}}]{haule2010dynamical}%
  \BibitemOpen
  \bibfield  {author} {\bibinfo {author} {\bibfnamefont {K.}~\bibnamefont
  {Haule}}, \bibinfo {author} {\bibfnamefont {C.-H.}\ \bibnamefont {Yee}}, \
  and\ \bibinfo {author} {\bibfnamefont {K.}~\bibnamefont {Kim}},\ }\href@noop
  {} {\bibfield  {journal} {\bibinfo  {journal} {Physical Review B}\ }\textbf
  {\bibinfo {volume} {81}},\ \bibinfo {pages} {195107} (\bibinfo {year}
  {2010})}\BibitemShut {NoStop}%
\bibitem [{\citenamefont {Haule}\ and\ \citenamefont
  {Birol}(2015)}]{haule2015free}%
  \BibitemOpen
  \bibfield  {author} {\bibinfo {author} {\bibfnamefont {K.}~\bibnamefont
  {Haule}}\ and\ \bibinfo {author} {\bibfnamefont {T.}~\bibnamefont {Birol}},\
  }\href@noop {} {\bibfield  {journal} {\bibinfo  {journal} {Physical Review
  Letters}\ }\textbf {\bibinfo {volume} {115}},\ \bibinfo {pages} {256402}
  (\bibinfo {year} {2015})}\BibitemShut {NoStop}%
\bibitem [{\citenamefont {Zhang}\ \emph {et~al.}(2016)\citenamefont {Zhang},
  \citenamefont {Cohen},\ and\ \citenamefont {Haule}}]{zhang2016retraction}%
  \BibitemOpen
  \bibfield  {author} {\bibinfo {author} {\bibfnamefont {P.}~\bibnamefont
  {Zhang}}, \bibinfo {author} {\bibfnamefont {R.}~\bibnamefont {Cohen}}, \ and\
  \bibinfo {author} {\bibfnamefont {K.}~\bibnamefont {Haule}},\ }\href@noop {}
  {\bibfield  {journal} {\bibinfo  {journal} {Nature}\ }\textbf {\bibinfo
  {volume} {536}},\ \bibinfo {pages} {112} (\bibinfo {year}
  {2016})}\BibitemShut {NoStop}%
\bibitem [{\citenamefont {Min{\'a}r}\ \emph {et~al.}(2005)\citenamefont
  {Min{\'a}r}, \citenamefont {Chioncel}, \citenamefont {Perlov}, \citenamefont
  {Ebert}, \citenamefont {Katsnelson},\ and\ \citenamefont
  {Lichtenstein}}]{minar2005multiple}%
  \BibitemOpen
  \bibfield  {author} {\bibinfo {author} {\bibfnamefont {J.}~\bibnamefont
  {Min{\'a}r}}, \bibinfo {author} {\bibfnamefont {L.}~\bibnamefont {Chioncel}},
  \bibinfo {author} {\bibfnamefont {A.}~\bibnamefont {Perlov}}, \bibinfo
  {author} {\bibfnamefont {H.}~\bibnamefont {Ebert}}, \bibinfo {author}
  {\bibfnamefont {M.}~\bibnamefont {Katsnelson}}, \ and\ \bibinfo {author}
  {\bibfnamefont {A.}~\bibnamefont {Lichtenstein}},\ }\href@noop {} {\bibfield
  {journal} {\bibinfo  {journal} {Physical Review B}\ }\textbf {\bibinfo
  {volume} {72}},\ \bibinfo {pages} {045125} (\bibinfo {year}
  {2005})}\BibitemShut {NoStop}%
\bibitem [{\citenamefont {Pourovskii}\ \emph {et~al.}(2005)\citenamefont
  {Pourovskii}, \citenamefont {Katsnelson},\ and\ \citenamefont
  {Lichtenstein}}]{pourovskii2005correlation}%
  \BibitemOpen
  \bibfield  {author} {\bibinfo {author} {\bibfnamefont {L.}~\bibnamefont
  {Pourovskii}}, \bibinfo {author} {\bibfnamefont {M.}~\bibnamefont
  {Katsnelson}}, \ and\ \bibinfo {author} {\bibfnamefont {A.}~\bibnamefont
  {Lichtenstein}},\ }\href@noop {} {\bibfield  {journal} {\bibinfo  {journal}
  {Physical Review B}\ }\textbf {\bibinfo {volume} {72}},\ \bibinfo {pages}
  {115106} (\bibinfo {year} {2005})}\BibitemShut {NoStop}%
\bibitem [{\citenamefont {Mandal}\ \emph {et~al.}(2014)\citenamefont {Mandal},
  \citenamefont {Cohen},\ and\ \citenamefont {Haule}}]{mandal2014strong}%
  \BibitemOpen
  \bibfield  {author} {\bibinfo {author} {\bibfnamefont {S.}~\bibnamefont
  {Mandal}}, \bibinfo {author} {\bibfnamefont {R.~E.}\ \bibnamefont {Cohen}}, \
  and\ \bibinfo {author} {\bibfnamefont {K.}~\bibnamefont {Haule}},\
  }\href@noop {} {\bibfield  {journal} {\bibinfo  {journal} {Physical Review
  B}\ }\textbf {\bibinfo {volume} {89}},\ \bibinfo {pages} {220502} (\bibinfo
  {year} {2014})}\BibitemShut {NoStop}%
\bibitem [{\citenamefont {Gerber}\ \emph {et~al.}(2017)\citenamefont {Gerber},
  \citenamefont {Yang}, \citenamefont {Zhu}, \citenamefont {Soifer},
  \citenamefont {Sobota}, \citenamefont {Rebec}, \citenamefont {Lee},
  \citenamefont {Jia}, \citenamefont {Moritz}, \citenamefont {Jia} \emph
  {et~al.}}]{gerber2017femtosecond}%
  \BibitemOpen
  \bibfield  {author} {\bibinfo {author} {\bibfnamefont {S.}~\bibnamefont
  {Gerber}}, \bibinfo {author} {\bibfnamefont {S.-L.}\ \bibnamefont {Yang}},
  \bibinfo {author} {\bibfnamefont {D.}~\bibnamefont {Zhu}}, \bibinfo {author}
  {\bibfnamefont {H.}~\bibnamefont {Soifer}}, \bibinfo {author} {\bibfnamefont
  {J.}~\bibnamefont {Sobota}}, \bibinfo {author} {\bibfnamefont
  {S.}~\bibnamefont {Rebec}}, \bibinfo {author} {\bibfnamefont
  {J.}~\bibnamefont {Lee}}, \bibinfo {author} {\bibfnamefont {T.}~\bibnamefont
  {Jia}}, \bibinfo {author} {\bibfnamefont {B.}~\bibnamefont {Moritz}},
  \bibinfo {author} {\bibfnamefont {C.}~\bibnamefont {Jia}},  \emph {et~al.},\
  }\href@noop {} {\bibfield  {journal} {\bibinfo  {journal} {Science}\ }\textbf
  {\bibinfo {volume} {357}},\ \bibinfo {pages} {71} (\bibinfo {year}
  {2017})}\BibitemShut {NoStop}%
\bibitem [{\citenamefont {Hausoel}\ \emph {et~al.}(2017)\citenamefont
  {Hausoel}, \citenamefont {Karolak}, \citenamefont
  {{\c{S}}a{\c{s}}$\iota$o{\u{g}}lu}, \citenamefont {Lichtenstein},
  \citenamefont {Held}, \citenamefont {Katanin}, \citenamefont {Toschi},\ and\
  \citenamefont {Sangiovanni}}]{hausoel2017local}%
  \BibitemOpen
  \bibfield  {author} {\bibinfo {author} {\bibfnamefont {A.}~\bibnamefont
  {Hausoel}}, \bibinfo {author} {\bibfnamefont {M.}~\bibnamefont {Karolak}},
  \bibinfo {author} {\bibfnamefont {E.}~\bibnamefont
  {{\c{S}}a{\c{s}}$\iota$o{\u{g}}lu}}, \bibinfo {author} {\bibfnamefont
  {A.}~\bibnamefont {Lichtenstein}}, \bibinfo {author} {\bibfnamefont
  {K.}~\bibnamefont {Held}}, \bibinfo {author} {\bibfnamefont {A.}~\bibnamefont
  {Katanin}}, \bibinfo {author} {\bibfnamefont {A.}~\bibnamefont {Toschi}}, \
  and\ \bibinfo {author} {\bibfnamefont {G.}~\bibnamefont {Sangiovanni}},\
  }\href@noop {} {\bibfield  {journal} {\bibinfo  {journal} {Nature
  communications}\ }\textbf {\bibinfo {volume} {8}},\ \bibinfo {pages} {16062}
  (\bibinfo {year} {2017})}\BibitemShut {NoStop}%
\bibitem [{\citenamefont {Gurvitch}(1981)}]{gurvitch1981ioffe}%
  \BibitemOpen
  \bibfield  {author} {\bibinfo {author} {\bibfnamefont {M.}~\bibnamefont
  {Gurvitch}},\ }\href@noop {} {\bibfield  {journal} {\bibinfo  {journal}
  {Physical Review B}\ }\textbf {\bibinfo {volume} {24}},\ \bibinfo {pages}
  {7404} (\bibinfo {year} {1981})}\BibitemShut {NoStop}%
\bibitem [{\citenamefont {Gunnarsson}\ \emph {et~al.}(2003)\citenamefont
  {Gunnarsson}, \citenamefont {Calandra},\ and\ \citenamefont
  {Han}}]{gunnarsson2003colloquium}%
  \BibitemOpen
  \bibfield  {author} {\bibinfo {author} {\bibfnamefont {O.}~\bibnamefont
  {Gunnarsson}}, \bibinfo {author} {\bibfnamefont {M.}~\bibnamefont
  {Calandra}}, \ and\ \bibinfo {author} {\bibfnamefont {J.}~\bibnamefont
  {Han}},\ }\href@noop {} {\bibfield  {journal} {\bibinfo  {journal} {Reviews
  of Modern Physics}\ }\textbf {\bibinfo {volume} {75}},\ \bibinfo {pages}
  {1085} (\bibinfo {year} {2003})}\BibitemShut {NoStop}%
\bibitem [{\citenamefont {Rizzo}\ \emph {et~al.}(2005)\citenamefont {Rizzo},
  \citenamefont {Cappelluti},\ and\ \citenamefont
  {Pietronero}}]{rizzo2005transport}%
  \BibitemOpen
  \bibfield  {author} {\bibinfo {author} {\bibfnamefont {F.}~\bibnamefont
  {Rizzo}}, \bibinfo {author} {\bibfnamefont {E.}~\bibnamefont {Cappelluti}}, \
  and\ \bibinfo {author} {\bibfnamefont {L.}~\bibnamefont {Pietronero}},\
  }\href@noop {} {\bibfield  {journal} {\bibinfo  {journal} {Physical Review
  B}\ }\textbf {\bibinfo {volume} {72}},\ \bibinfo {pages} {155113} (\bibinfo
  {year} {2005})}\BibitemShut {NoStop}%
\bibitem [{\citenamefont {Desjarlais}\ \emph {et~al.}(2017)\citenamefont
  {Desjarlais}, \citenamefont {Scullard}, \citenamefont {Benedict},
  \citenamefont {Whitley},\ and\ \citenamefont
  {Redmer}}]{desjarlais2017density}%
  \BibitemOpen
  \bibfield  {author} {\bibinfo {author} {\bibfnamefont {M.~P.}\ \bibnamefont
  {Desjarlais}}, \bibinfo {author} {\bibfnamefont {C.~R.}\ \bibnamefont
  {Scullard}}, \bibinfo {author} {\bibfnamefont {L.~X.}\ \bibnamefont
  {Benedict}}, \bibinfo {author} {\bibfnamefont {H.~D.}\ \bibnamefont
  {Whitley}}, \ and\ \bibinfo {author} {\bibfnamefont {R.}~\bibnamefont
  {Redmer}},\ }\href@noop {} {\bibfield  {journal} {\bibinfo  {journal}
  {Physical Review E}\ }\textbf {\bibinfo {volume} {95}},\ \bibinfo {pages}
  {033203} (\bibinfo {year} {2017})}\BibitemShut {NoStop}%
\bibitem [{\citenamefont {Pourovskii}\ \emph {et~al.}(2017)\citenamefont
  {Pourovskii}, \citenamefont {Mravlje}, \citenamefont {Georges}, \citenamefont
  {Simak},\ and\ \citenamefont {Abrikosov}}]{pourovskii2017electron}%
  \BibitemOpen
  \bibfield  {author} {\bibinfo {author} {\bibfnamefont {L.}~\bibnamefont
  {Pourovskii}}, \bibinfo {author} {\bibfnamefont {J.}~\bibnamefont {Mravlje}},
  \bibinfo {author} {\bibfnamefont {A.}~\bibnamefont {Georges}}, \bibinfo
  {author} {\bibfnamefont {S.}~\bibnamefont {Simak}}, \ and\ \bibinfo {author}
  {\bibfnamefont {I.}~\bibnamefont {Abrikosov}},\ }\href@noop {} {\bibfield
  {journal} {\bibinfo  {journal} {New Journal of Physics}\ }\textbf {\bibinfo
  {volume} {19}},\ \bibinfo {pages} {073022} (\bibinfo {year}
  {2017})}\BibitemShut {NoStop}%
\bibitem [{\citenamefont {Bi}\ \emph {et~al.}(2002)\citenamefont {Bi},
  \citenamefont {Tan},\ and\ \citenamefont {Jing}}]{bi2002electrical}%
  \BibitemOpen
  \bibfield  {author} {\bibinfo {author} {\bibfnamefont {Y.}~\bibnamefont
  {Bi}}, \bibinfo {author} {\bibfnamefont {H.}~\bibnamefont {Tan}}, \ and\
  \bibinfo {author} {\bibfnamefont {F.}~\bibnamefont {Jing}},\ }\href@noop {}
  {\bibfield  {journal} {\bibinfo  {journal} {Journal of Physics: Condensed
  Matter}\ }\textbf {\bibinfo {volume} {14}},\ \bibinfo {pages} {10849}
  (\bibinfo {year} {2002})}\BibitemShut {NoStop}%
\bibitem [{\citenamefont {Madsen}\ and\ \citenamefont
  {Singh}(2006)}]{madsen2006boltztrap}%
  \BibitemOpen
  \bibfield  {author} {\bibinfo {author} {\bibfnamefont {G.~K.}\ \bibnamefont
  {Madsen}}\ and\ \bibinfo {author} {\bibfnamefont {D.~J.}\ \bibnamefont
  {Singh}},\ }\href@noop {} {\bibfield  {journal} {\bibinfo  {journal}
  {Computer Physics Communications}\ }\textbf {\bibinfo {volume} {175}},\
  \bibinfo {pages} {67} (\bibinfo {year} {2006})}\BibitemShut {NoStop}%
\bibitem [{\citenamefont {Pozzo}\ and\ \citenamefont
  {Alf{\`e}}(2016)}]{pozzo2016saturation}%
  \BibitemOpen
  \bibfield  {author} {\bibinfo {author} {\bibfnamefont {M.}~\bibnamefont
  {Pozzo}}\ and\ \bibinfo {author} {\bibfnamefont {D.}~\bibnamefont
  {Alf{\`e}}},\ }\href@noop {} {\bibfield  {journal} {\bibinfo  {journal}
  {SpringerPlus}\ }\textbf {\bibinfo {volume} {5}},\ \bibinfo {pages} {256}
  (\bibinfo {year} {2016})}\BibitemShut {NoStop}%
\bibitem [{\citenamefont {Stacey}\ and\ \citenamefont
  {Anderson}(2001)}]{stacey2001electrical}%
  \BibitemOpen
  \bibfield  {author} {\bibinfo {author} {\bibfnamefont {F.~D.}\ \bibnamefont
  {Stacey}}\ and\ \bibinfo {author} {\bibfnamefont {O.~L.}\ \bibnamefont
  {Anderson}},\ }\href@noop {} {\bibfield  {journal} {\bibinfo  {journal}
  {Physics of the Earth and Planetary Interiors}\ }\textbf {\bibinfo {volume}
  {124}},\ \bibinfo {pages} {153} (\bibinfo {year} {2001})}\BibitemShut
  {NoStop}%
\bibitem [{\citenamefont {Pozzo}\ \emph {et~al.}(2014)\citenamefont {Pozzo},
  \citenamefont {Davies}, \citenamefont {Gubbins},\ and\ \citenamefont
  {Alf{\`e}}}]{pozzo2014thermal}%
  \BibitemOpen
  \bibfield  {author} {\bibinfo {author} {\bibfnamefont {M.}~\bibnamefont
  {Pozzo}}, \bibinfo {author} {\bibfnamefont {C.}~\bibnamefont {Davies}},
  \bibinfo {author} {\bibfnamefont {D.}~\bibnamefont {Gubbins}}, \ and\
  \bibinfo {author} {\bibfnamefont {D.}~\bibnamefont {Alf{\`e}}},\ }\href@noop
  {} {\bibfield  {journal} {\bibinfo  {journal} {Earth and Planetary Science
  Letters}\ }\textbf {\bibinfo {volume} {393}},\ \bibinfo {pages} {159}
  (\bibinfo {year} {2014})}\BibitemShut {NoStop}%
\bibitem [{\citenamefont {Lay}\ \emph {et~al.}(2008)\citenamefont {Lay},
  \citenamefont {Hernlund},\ and\ \citenamefont {Buffett}}]{lay2008core}%
  \BibitemOpen
  \bibfield  {author} {\bibinfo {author} {\bibfnamefont {T.}~\bibnamefont
  {Lay}}, \bibinfo {author} {\bibfnamefont {J.}~\bibnamefont {Hernlund}}, \
  and\ \bibinfo {author} {\bibfnamefont {B.~A.}\ \bibnamefont {Buffett}},\
  }\href@noop {} {\bibfield  {journal} {\bibinfo  {journal} {Nature
  geoscience}\ }\textbf {\bibinfo {volume} {1}},\ \bibinfo {pages} {25}
  (\bibinfo {year} {2008})}\BibitemShut {NoStop}%
\bibitem [{\citenamefont {Wu}\ \emph {et~al.}(2011)\citenamefont {Wu},
  \citenamefont {Driscoll},\ and\ \citenamefont {Olson}}]{wu2011statistical}%
  \BibitemOpen
  \bibfield  {author} {\bibinfo {author} {\bibfnamefont {B.}~\bibnamefont
  {Wu}}, \bibinfo {author} {\bibfnamefont {P.}~\bibnamefont {Driscoll}}, \ and\
  \bibinfo {author} {\bibfnamefont {P.}~\bibnamefont {Olson}},\ }\href@noop {}
  {\bibfield  {journal} {\bibinfo  {journal} {Journal of Geophysical Research:
  Solid Earth}\ }\textbf {\bibinfo {volume} {116}},\ \bibinfo {pages} {B12112}
  (\bibinfo {year} {2011})}\BibitemShut {NoStop}%
\bibitem [{\citenamefont {Belonoshko}\ \emph {et~al.}(2017)\citenamefont
  {Belonoshko}, \citenamefont {Lukinov}, \citenamefont {Fu}, \citenamefont
  {Zhao}, \citenamefont {Davis},\ and\ \citenamefont
  {Simak}}]{belonoshko2017stabilization}%
  \BibitemOpen
  \bibfield  {author} {\bibinfo {author} {\bibfnamefont {A.~B.}\ \bibnamefont
  {Belonoshko}}, \bibinfo {author} {\bibfnamefont {T.}~\bibnamefont {Lukinov}},
  \bibinfo {author} {\bibfnamefont {J.}~\bibnamefont {Fu}}, \bibinfo {author}
  {\bibfnamefont {J.}~\bibnamefont {Zhao}}, \bibinfo {author} {\bibfnamefont
  {S.}~\bibnamefont {Davis}}, \ and\ \bibinfo {author} {\bibfnamefont {S.~I.}\
  \bibnamefont {Simak}},\ }\href@noop {} {\bibfield  {journal} {\bibinfo
  {journal} {Nature Geoscience}\ }\textbf {\bibinfo {volume} {10}},\ \bibinfo
  {pages} {312} (\bibinfo {year} {2017})}\BibitemShut {NoStop}%
\bibitem [{\citenamefont {Vocadlo}\ \emph {et~al.}(2008)\citenamefont
  {Vocadlo}, \citenamefont {Wood}, \citenamefont {Gillan}, \citenamefont
  {Brodholt}, \citenamefont {Dobson}, \citenamefont {Price},\ and\
  \citenamefont {Alfe}}]{Vocadlo2008}%
  \BibitemOpen
  \bibfield  {author} {\bibinfo {author} {\bibfnamefont {L.}~\bibnamefont
  {Vocadlo}}, \bibinfo {author} {\bibfnamefont {I.~G.}\ \bibnamefont {Wood}},
  \bibinfo {author} {\bibfnamefont {M.~J.}\ \bibnamefont {Gillan}}, \bibinfo
  {author} {\bibfnamefont {J.}~\bibnamefont {Brodholt}}, \bibinfo {author}
  {\bibfnamefont {D.~P.}\ \bibnamefont {Dobson}}, \bibinfo {author}
  {\bibfnamefont {G.~D.}\ \bibnamefont {Price}}, \ and\ \bibinfo {author}
  {\bibfnamefont {D.}~\bibnamefont {Alfe}},\ }\href@noop {} {\bibfield
  {journal} {\bibinfo  {journal} {Physics of the Earth and Planetary
  Interiors}\ }\textbf {\bibinfo {volume} {170}},\ \bibinfo {pages} {52}
  (\bibinfo {year} {2008})}\BibitemShut {NoStop}%
\bibitem [{\citenamefont {Driscoll}\ and\ \citenamefont
  {Bercovici}(2014)}]{Driscoll2014}%
  \BibitemOpen
  \bibfield  {author} {\bibinfo {author} {\bibfnamefont {P.}~\bibnamefont
  {Driscoll}}\ and\ \bibinfo {author} {\bibfnamefont {D.}~\bibnamefont
  {Bercovici}},\ }\href {\doibase 10.1016/j.pepi.2014.08.004} {\bibfield
  {journal} {\bibinfo  {journal} {Physics of the Earth and Planetary
  Interiors}\ }\textbf {\bibinfo {volume} {236}},\ \bibinfo {pages} {36}
  (\bibinfo {year} {2014})}\BibitemShut {NoStop}%
\bibitem [{\citenamefont {Goedecker}\ \emph {et~al.}(1996)\citenamefont
  {Goedecker}, \citenamefont {Teter},\ and\ \citenamefont
  {Hutter}}]{goedecker1996separable}%
  \BibitemOpen
  \bibfield  {author} {\bibinfo {author} {\bibfnamefont {S.}~\bibnamefont
  {Goedecker}}, \bibinfo {author} {\bibfnamefont {M.}~\bibnamefont {Teter}}, \
  and\ \bibinfo {author} {\bibfnamefont {J.}~\bibnamefont {Hutter}},\
  }\href@noop {} {\bibfield  {journal} {\bibinfo  {journal} {Physical Review
  B}\ }\textbf {\bibinfo {volume} {54}},\ \bibinfo {pages} {1703} (\bibinfo
  {year} {1996})}\BibitemShut {NoStop}%
\bibitem [{\citenamefont {Perdew}\ \emph {et~al.}(1996)\citenamefont {Perdew},
  \citenamefont {Burke},\ and\ \citenamefont
  {Ernzerhof}}]{perdew1996generalized}%
  \BibitemOpen
  \bibfield  {author} {\bibinfo {author} {\bibfnamefont {J.~P.}\ \bibnamefont
  {Perdew}}, \bibinfo {author} {\bibfnamefont {K.}~\bibnamefont {Burke}}, \
  and\ \bibinfo {author} {\bibfnamefont {M.}~\bibnamefont {Ernzerhof}},\
  }\href@noop {} {\bibfield  {journal} {\bibinfo  {journal} {Physical review
  letters}\ }\textbf {\bibinfo {volume} {77}},\ \bibinfo {pages} {3865}
  (\bibinfo {year} {1996})}\BibitemShut {NoStop}%
\bibitem [{\citenamefont {Schwarz}\ \emph {et~al.}(2002)\citenamefont
  {Schwarz}, \citenamefont {Blaha},\ and\ \citenamefont
  {Madsen}}]{schwarz2002electronic}%
  \BibitemOpen
  \bibfield  {author} {\bibinfo {author} {\bibfnamefont {K.}~\bibnamefont
  {Schwarz}}, \bibinfo {author} {\bibfnamefont {P.}~\bibnamefont {Blaha}}, \
  and\ \bibinfo {author} {\bibfnamefont {G.~K.}\ \bibnamefont {Madsen}},\
  }\href@noop {} {\bibfield  {journal} {\bibinfo  {journal} {Computer Physics
  Communications}\ }\textbf {\bibinfo {volume} {147}},\ \bibinfo {pages} {71}
  (\bibinfo {year} {2002})}\BibitemShut {NoStop}%
\bibitem [{\citenamefont {Min{\'a}r}(2011)}]{minar2011correlation}%
  \BibitemOpen
  \bibfield  {author} {\bibinfo {author} {\bibfnamefont {J.}~\bibnamefont
  {Min{\'a}r}},\ }\href@noop {} {\bibfield  {journal} {\bibinfo  {journal}
  {Journal of Physics: Condensed Matter}\ }\textbf {\bibinfo {volume} {23}},\
  \bibinfo {pages} {253201} (\bibinfo {year} {2011})}\BibitemShut {NoStop}%
\bibitem [{\citenamefont {Sch{\"o}tt}\ \emph {et~al.}(2016)\citenamefont
  {Sch{\"o}tt}, \citenamefont {Locht}, \citenamefont {Lundin}, \citenamefont
  {Gr{\aa}n{\"a}s}, \citenamefont {Eriksson},\ and\ \citenamefont
  {Di~Marco}}]{schott2016analytic}%
  \BibitemOpen
  \bibfield  {author} {\bibinfo {author} {\bibfnamefont {J.}~\bibnamefont
  {Sch{\"o}tt}}, \bibinfo {author} {\bibfnamefont {I.~L.}\ \bibnamefont
  {Locht}}, \bibinfo {author} {\bibfnamefont {E.}~\bibnamefont {Lundin}},
  \bibinfo {author} {\bibfnamefont {O.}~\bibnamefont {Gr{\aa}n{\"a}s}},
  \bibinfo {author} {\bibfnamefont {O.}~\bibnamefont {Eriksson}}, \ and\
  \bibinfo {author} {\bibfnamefont {I.}~\bibnamefont {Di~Marco}},\ }\href@noop
  {} {\bibfield  {journal} {\bibinfo  {journal} {Physical Review B}\ }\textbf
  {\bibinfo {volume} {93}},\ \bibinfo {pages} {075104} (\bibinfo {year}
  {2016})}\BibitemShut {NoStop}%
\bibitem [{\citenamefont {Nozi{\`e}res}\ and\ \citenamefont
  {Pines}(1999)}]{nozieres1999theory}%
  \BibitemOpen
  \bibfield  {author} {\bibinfo {author} {\bibfnamefont {P.}~\bibnamefont
  {Nozi{\`e}res}}\ and\ \bibinfo {author} {\bibfnamefont {D.}~\bibnamefont
  {Pines}},\ }\href@noop {} {\emph {\bibinfo {title} {Theory of quantum
  liquids}}}\ (\bibinfo  {publisher} {Westview Press},\ \bibinfo {year}
  {1999})\BibitemShut {NoStop}%
\bibitem [{\citenamefont {Herring}(1967{\natexlab{a}})}]{herring1967simple}%
  \BibitemOpen
  \bibfield  {author} {\bibinfo {author} {\bibfnamefont {C.}~\bibnamefont
  {Herring}},\ }\href@noop {} {\bibfield  {journal} {\bibinfo  {journal}
  {Physical Review Letters}\ }\textbf {\bibinfo {volume} {19}},\ \bibinfo
  {pages} {167} (\bibinfo {year} {1967}{\natexlab{a}})}\BibitemShut {NoStop}%
\bibitem [{\citenamefont {Herring}(1967{\natexlab{b}})}]{herring1967simple2}%
  \BibitemOpen
  \bibfield  {author} {\bibinfo {author} {\bibfnamefont {C.}~\bibnamefont
  {Herring}},\ }\href@noop {} {\bibfield  {journal} {\bibinfo  {journal}
  {Physical Review Letters}\ }\textbf {\bibinfo {volume} {19}},\ \bibinfo
  {pages} {684} (\bibinfo {year} {1967}{\natexlab{b}})}\BibitemShut {NoStop}%
\bibitem [{\citenamefont {Chubukov}\ and\ \citenamefont
  {Maslov}(2012)}]{chubukov2012first}%
  \BibitemOpen
  \bibfield  {author} {\bibinfo {author} {\bibfnamefont {A.~V.}\ \bibnamefont
  {Chubukov}}\ and\ \bibinfo {author} {\bibfnamefont {D.~L.}\ \bibnamefont
  {Maslov}},\ }\href@noop {} {\bibfield  {journal} {\bibinfo  {journal}
  {Physical Review B}\ }\textbf {\bibinfo {volume} {86}},\ \bibinfo {pages}
  {155136} (\bibinfo {year} {2012})}\BibitemShut {NoStop}%
\bibitem [{\citenamefont {Milchberg}\ \emph {et~al.}(1988)\citenamefont
  {Milchberg}, \citenamefont {Freeman}, \citenamefont {Davey},\ and\
  \citenamefont {More}}]{milchberg1988resistivity}%
  \BibitemOpen
  \bibfield  {author} {\bibinfo {author} {\bibfnamefont {H.}~\bibnamefont
  {Milchberg}}, \bibinfo {author} {\bibfnamefont {R.}~\bibnamefont {Freeman}},
  \bibinfo {author} {\bibfnamefont {S.}~\bibnamefont {Davey}}, \ and\ \bibinfo
  {author} {\bibfnamefont {R.}~\bibnamefont {More}},\ }\href@noop {} {\bibfield
   {journal} {\bibinfo  {journal} {Physical review letters}\ }\textbf {\bibinfo
  {volume} {61}},\ \bibinfo {pages} {2364} (\bibinfo {year}
  {1988})}\BibitemShut {NoStop}%
\bibitem [{\citenamefont {Abraham}\ and\ \citenamefont
  {Deviot}(1972)}]{abraham1972resistivite}%
  \BibitemOpen
  \bibfield  {author} {\bibinfo {author} {\bibfnamefont {J.}~\bibnamefont
  {Abraham}}\ and\ \bibinfo {author} {\bibfnamefont {B.}~\bibnamefont
  {Deviot}},\ }\href@noop {} {\bibfield  {journal} {\bibinfo  {journal}
  {Journal of the Less Common Metals}\ }\textbf {\bibinfo {volume} {29}},\
  \bibinfo {pages} {311} (\bibinfo {year} {1972})}\BibitemShut {NoStop}%
\bibitem [{\citenamefont {Giannozzi}\ \emph {et~al.}(2009)\citenamefont
  {Giannozzi}, \citenamefont {Baroni}, \citenamefont {Bonini}, \citenamefont
  {Calandra}, \citenamefont {Car}, \citenamefont {Cavazzoni}, \citenamefont
  {Ceresoli}, \citenamefont {Chiarotti}, \citenamefont {Cococcioni},
  \citenamefont {Dabo} \emph {et~al.}}]{giannozzi2009quantum}%
  \BibitemOpen
  \bibfield  {author} {\bibinfo {author} {\bibfnamefont {P.}~\bibnamefont
  {Giannozzi}}, \bibinfo {author} {\bibfnamefont {S.}~\bibnamefont {Baroni}},
  \bibinfo {author} {\bibfnamefont {N.}~\bibnamefont {Bonini}}, \bibinfo
  {author} {\bibfnamefont {M.}~\bibnamefont {Calandra}}, \bibinfo {author}
  {\bibfnamefont {R.}~\bibnamefont {Car}}, \bibinfo {author} {\bibfnamefont
  {C.}~\bibnamefont {Cavazzoni}}, \bibinfo {author} {\bibfnamefont
  {D.}~\bibnamefont {Ceresoli}}, \bibinfo {author} {\bibfnamefont {G.~L.}\
  \bibnamefont {Chiarotti}}, \bibinfo {author} {\bibfnamefont {M.}~\bibnamefont
  {Cococcioni}}, \bibinfo {author} {\bibfnamefont {I.}~\bibnamefont {Dabo}},
  \emph {et~al.},\ }\href@noop {} {\bibfield  {journal} {\bibinfo  {journal}
  {Journal of physics: Condensed matter}\ }\textbf {\bibinfo {volume} {21}},\
  \bibinfo {pages} {395502} (\bibinfo {year} {2009})}\BibitemShut {NoStop}%
\end{thebibliography}%


\clearpage

\onecolumngrid
\begin{center}
  \textbf{\large Thermal conductivity and electrical resistivity of solid iron at Earth's core conditions\\from first principles - Supplemental Material}\\[.2cm]
  Junqing Xu,$^{1}$ Peng Zhang,$^{2}$ K. Haule,$^{3}$ Jan Minar,$^{4}$ Sebastian Wimmer,$^{5}$ Hubert Ebert,$^{5}$ and R. E. Cohen,$^{1,6}$ \\[.1cm]
  {\itshape ${}^1$Department of Earth and Environmental Sciences, LMU Munich, Theresienstrasse 41, 80333 Munich, Germany\\
  ${}^2$Department of Physics, Xi'an Jiaotong University, Xi'an, Shanxi, 710049, P. R. China\\
  ${}^3$Department of Physics, Rutgers University, Piscataway, New Jersey 08854, USA\\
  ${}^4$University of West Bohemia, New Technologies - Research Centre, Pllsen, Czech Republic\\
  ${}^5$Department Chemie,  Physikalische Chemie, University of Munich, D-81377 Munich, Germany\\
  ${}^6$Extreme Materials Initiative, Geophysical Laboratory, Carnegie Institution for Science, Washington, DC 20015-1305, USA\\}
(Dated: \today)\\[1cm]
\end{center}
\twocolumngrid

\setcounter{equation}{0}
\setcounter{figure}{0}
\setcounter{table}{0}
\setcounter{page}{1}
\renewcommand{\theequation}{S\arabic{equation}}
\renewcommand{\thefigure}{S\arabic{figure}}

\maketitle

\section{Equation of state}
We use the experimental equation of state (EOS) of Ref. \cite{dewaele2006quasihydrostatic} , which agrees well
with the theoretical EOS of Ref. \cite{sha2010first} at pressures above 50 GPa. Our calculations are all at fixed volume and temperature, so can be easily referred to any equation of state.
\section{Technical details of transport properties of hcp iron using DFPT}
We used the local density approximation (LDA) \cite{goedecker1996separable} , and also compared with the Perdew-Burke-Ernzerhof (PBE) functional \citep{perdew1996generalized}. We also compared different pseudopotentials. Our results were robust within 2\% at given volume and temperature. Single particle orbitals are occupied according to Fermi-Dirac statistics for each temperature.
{$\bf{k}$}-point grids 24$\times$24$\times$16 and ${\bf{q}}$-point grids 12$\times$12$\times$8 are used.
\section{Estimates of saturation resistivity}
We use the widely used criterion mean free path $l_{Boltzmann}\,=\,a$ to estimate saturation resistivity $\rho_{sat}$, i.e., Ioffe-Regel limit.
Here $a$ is the nearest-neighbour distance and is nearly the same as the lattice constant for hcp Fe.
We compute the minimum relaxation time $\tau_0\,=\,a/\bar{v_F}$, where $\bar{v_F}$ is the mean square root of Fermi velocity square.
$\rho_{sat}$ is obtained by replacing $\tau_{Boltzmann}$ by $\tau_0$.
The {\sc BoltzTraP} code \cite{madsen2006boltztrap} is used to compute $\sigma_{Boltzmann}$/$\tau_{Boltzmann}$ and $\bar{v_F}$ for the estimates of $\rho_{sat}$.
Our estimates of $\rho_{sat}$ of hcp Fe at Earth's outer and inner conditions are 155 and 143 $\mu\Omega$ cm, respectively.
To ensure the reliability of our method of estimating $\rho_{sat}$. We have applies it to two element metals - Al and Nb.
The resulting $\rho_{sat}$ of Al is 169 $\mu\Omega$ cm in agreement with the reported value in Ref. \cite{milchberg1988resistivity}.
The resulting $\rho_{sat}$ of Nb is 133 $\mu\Omega$ cm. Our calculated DFPT result of $\rho_{Boltzmann}$ of Nb is about 56 $\mu\Omega$ cm at 1000 K,
so that $\rho_{e-ph}$ of Nb is about 39 $\mu\Omega$ cm in good agreement with experimental data \cite{abraham1972resistivite}.

\section{Electrical resistivity of hcp iron using KKR-CPA}
In the KKR-CPA, the electrical resistivity due to electron-phonon scattering is computed within the Born-Oppenheimer approximation, using the Kubo-Greenwood formula for alloy \cite{ebert2015calculating}.
The system at finite temperature as
an alloy whose components are generated by displacements of atoms according to their quasiharmonic (or anharmonic) probability distributions.
To make use of the single-site CPA for the alloy, the displacement of each atom is assumed independent of its neighbors.
The displacement distribution can be either from lattice dynamics or molecular dynamics. In this work, we distribute the displacement
according to lattice dynamics using the phonon frequencies and eigenvectors of hcp Fe calculated by ABINIT based on DFPT \cite{gonze2016}.
We tested the effects of anisotropy of the displacement distribution and found very little impact even at core temperatures for hcp Fe. So we distribute the radial displacements according to the computed quasiharmonic phonon density of state.

We used a maximum angular momentum $l_{max}$ of 3, but tested  $l_{max}$ until 5 and found only tiny differences.
The integrals for the Kubo-Greenwood formula are done on 100,000 k points and energies [-10\,$k_B\,T$,10\,$k_B\,T$] around the chemical potential.

Although the displacement distribution is determined in a quasiharmonic approximation, resistivity saturation is indeed observed.
This is because unlike DFPT, KKR-CPA is non-perturbative.

\section{Transport calculation using WIEN2K-DMFT}

The Hubbard parameter $U$ and the Hund's coupling $J$ are 5 eV and 0.943 eV, respectively.
The double counting energy $E_{dc}$ is calculated from the fully localized limit
formula $E_{dc}\,=\,U(n^0_{cor}-0.5)-0.5J(n^0_{cor}-1)$,
where $n^0_{cor}$ is nominal electron occupancy of the correlated atom and is set to 7.
The continuous time quantum Monte Carlo (CTQMC) impurity solver \cite{haule2007quantum,gull2011continuous}
is used to obtain the self-energy on the imaginary-frequency axis, $\Sigma(i\omega_n)$.
To obtain the self-energy along the real-frequency axis, $\Sigma(\varepsilon)$,
we use maximum entropy analytic continuation, which was compared and tested against Pad{\'e} and singular value decomposition methods,
giving identical results. The lattice is solved with the correlated self-energy
using the all-electron LAPW (linearized augmented plane wave) WIEN2K code \cite{schwarz2002electronic}.\par

Neglecting the vertex corrections, the dc electrical resistivity $\rho_{e-e}$ and the thermal conductivity $\kappa_{e-e}$ can be computed using Kubo-Greenwood formula:
\begin{gather*}
L_{ij}  =  (-1)^{i+j}\frac{\pi e^2}{V} \sum_k \int d\varepsilon (-\frac{df}{d\varepsilon}) Tr[\rho_k(\varepsilon)v_k\rho_k(\varepsilon)v_k]\varepsilon^{i+j-2}, \\
\rho_{e-e}  =  \frac{1}{L_{11}}, \\
\kappa_{e-e}  =  \frac{1}{e^2 T}(L_{22} - \frac{L^2_{12}}{L_{11}}),
\end{gather*}
where $f$ is the Fermi-Dirac distribution function, $v$ is the velocity and $\rho_k$ is the spectral function at wave vector $k$ and is related to the Green's function $G$($\varepsilon$) by $\rho_k$ = $(G^\dag-G)/(2\pi i)$.
By solving the complex eigenvalue problems $H A^R = E A^R$ and $A^L H = E A^L$, where $H$ is the Hamiltonian with added self-energy along the real-energy axis, $\Sigma$($\varepsilon$), we obtain the eigenvalues $E$ and the eigenvectors $A^R$ and $A^L$, so that $G$($\varepsilon$) = $A^R$\textbf{1}/($\varepsilon$-$E$) $A^L$. Compared with the formulas of Ref. \cite{haule2010dynamical}, we go beyond the approximate form of $\Sigma(\varepsilon) \approx \Sigma(0) + ( 1 - Z^{-1} ) \varepsilon - i \varepsilon^2 B$ and use the full energy dependence of the self-energy, as obtained by analytic continuation.

The integrals are done on k-point grids 26$\times$26$\times$14
and energies [-10\,$k_B\,T$,10\,$k_B\,T$] around the chemical potential.

\section{Comparison with previous DMFT calculations}
Other DFT-DMFT calculations for hcp Fe claim $\rho_{e-e} \propto T^2$, and claim nearly perfect Fermi liquid behavior of solid hcp Fe at all temperatures \cite{pourovskii2017electron}.
It has been proven that the strong energy-dependence of the relaxation time $\tau_{e-e}$ of a Fermi liquid \cite{nozieres1999theory}
will lead to a constant Lorenz number for e-e scattering, $L_{FL}$, about 1.58$\times$10$^{-8}$\,W$\Omega$K$^{-2}$ \cite{herring1967simple,herring1967simple2},
strongly reduced from the conventional $L_0$ of 2.44$\times$10$^{-8}$\,W$\Omega$K$^{-2}$.
However, their data of $\rho_{e-e}$ actually look linear at high temperatures.
Further, our DMFT results show that (i) our calculated $\rho_{e-e}$ and scattering rate are linear in temperature (Fig. S1 and S3) at T $>$ 2000 K,
(ii) the strength of the energy-dependence of $\tau_{e-e}$ is actually weaker than that of a simple Fermi liquid at 6000 K (Fig. S2),
and the Lorenz number for e-e scattering ranges from 1.4 to 2.0$\times$10$^{-8}$\,W$\Omega$K$^{-2}$ from 2000 K to 6000 K.

$Pourovskii\ et\ al.$ also resort to the first-Matsubara-frequency rule of $Chubokov\ and\ Maslov$ \cite{chubukov2012first},
and claim Im$\Sigma(i \pi T)$ being proportional to $T$ is proof of a simple Fermi liquid.
However, the paper by $Chubokov\ and\ Maslov$ proves something quite different:
namely, that Im$\Sigma(i\pi k_B T)$ is proportional to T not only in the Fermi liquid regime,
but also in many systems far beyond the Fermi liquid,
in particular it holds for "local approximation" (in which the interaction can be approximated by its value for the initial and final fermionic states right on the Fermi surface),
and for the marginal Fermi liquid in which the scattering rate is linear in temperature.
Hence the linear behavior of the self-energy at the first Matsubara point is not firm evidence of a simple Fermi liquid.

We observe that our calculated Im$\Sigma(i\pi k_B T)$ are close to those obtained by $Pourovskii\ et\ al.$ at all temperatures
and Im$\Sigma(i\pi k_B T)$ is proportional to $T$.
However, our calculated scattering rates and $\rho_{e-e}$ are 40$\%$ higher than theirs at Earth's inner core conditions.
The differences are either due to different values of Im$\Sigma$ at other Matsubara frequencies
or due to the way of doing analytic continuation.
Since we have done analytic continuation by three methods (see the above section) and obtained similar Im$\Sigma(\varepsilon)$,
our analytic continuation is accurate and robust.

We stress here that we observe that the scattering rate is frequency dependent,
as pointed out in Ref. \cite{pourovskii2017electron},
but the frequency dependence of the scattering rate has stronger effects on the thermal conductivity
when properly used in the Kubo-Greenwood formula, as compared to the simpler Fermi-liquid type approximation.
We also point out that in contrast to Ref. \cite{pourovskii2017electron},
our calculation shows that hcp iron is not a simple Fermi liquid at the high temperatures of Earth's core,
even when the feedback effect of the e-ph scattering on electronic correlations is neglected.
As the e-ph scattering is stronger than the e-e scattering,
the Fermi liquid approximation cannot be expected to hold under these conditions in any case.

\section{Transport calculation using KKR-DMFT}

We also calculate the electrical resistivity due to electron-electron scattering by fully relativistic KKR-DMFT \cite{minar2005multiple} implemented in the SPR-KKR package \cite{ebert2011calculating}.
The SPTF (spin-polarized T matrix + FLEX (fluctuation exchange)) impurity solver \cite{pourovskii2005correlation} is used.
Compared with the CTQMC impurity solver, this solver is based on the second order perturbation theory and includes FLEX diagrams from the third order.
As it was shown previously \cite{minar2011correlation}, this solver is suitable to describe correlation effects for moderately correlated systems
as for example transition metals and their alloys. In contrast to the CTQMC, the SPTF self-energy is analytical function
without statistical noise so that the errors due to analytical continuations are reduced.
Considering the correlated orbitals are constructed in the Wigner-Seitz sphere instead of the Muffin-Tin sphere as in WIEN2K-DMFT calculations,
to obtain similar values of resistivity, we choose a smaller Hubbard parameter $U$, 4.0 eV, but the same $J$, 0.943 eV.
The analytical continuations of the bath Green's function and electronic self-energy are done using the Pad{\'e} approximation.
Since in KKR-DMFT the lattice problem is solved with correlated self-energy at each self-consistent iteration and considering the non-diagonal nature of bath Green's function matrices in KKR-DMFT method, Pad{\'e} continuations are done many times
and the stability of Pad{\'e} continuation becomes serious. We apply the averaging Pad{\'e} approximation scheme suggested by Sch{\"o}tt\ et\ al. \cite{schott2016analytic}
to improve the accuracy and the stability of analytical continuation.

Electrical resistivity is again calculated using the Kubo-Greenwood formula for $l_{max}$ 3.
The integrals for the Kubo-Greenwood formula are done on 100,000 k points and energies [-10\,$k_B\,T$,10\,$k_B\,T$] around the chemical potential.
The resulting resistivities are shown in Fig. S3. It can be seen that the resistivity is linear in temperature above 3000 K and in agreement with our WIEN2K-DMFT calculations.

\section{Electron-electron scattering}

In Fig. S1-S3, we show scattering rates, relaxation times and resistivity due to electron-electron scattering.

\section{DMFT calculations of liquid Fe}
We perform First-principles molecular dynamics (FPMD) simulations of liquid Fe using the QUANTUM ESPRESSO package \cite{giannozzi2009quantum}. The simulation supercell contains 128 atoms. We use the GBRV ultrasoft pseudopotential and the plane-wave cut-off energy is 40 Ry. The Brillourin zone is sampled at the $\Gamma$ point only. Simulations are performed in the $NVT$ ensemble for longer than 10 ps at atomic volume 56.5 bohr$^3$ and temperature of 4000 K. We employ KKR-DMFT to selected snapshots which are separated from each other by at least 1 ps
and find that melting increases the scattering rates around the chemical potential due to e-e scattering by 20\% to 40\% for all snapshots (See Fig. S4)
and decreases thermal conductivity by about 10\%. The number of snapshots is enough to converge the transport properties to better than 2\%.

\section{Comparisons with previous work}

In Table S1-S6, our theoretical results of electrical resistivity and thermal conductivity are compared with other theoretical results, experimental data and results of models based on experimental data,
for three kinds of conditions - about 100 GPa and 2000 K, Core-Mantle Boundary conditions and Inner Core Boundary conditions.



\begin{figure*}
  \centering
  \noindent\includegraphics[width=1.0\columnwidth]{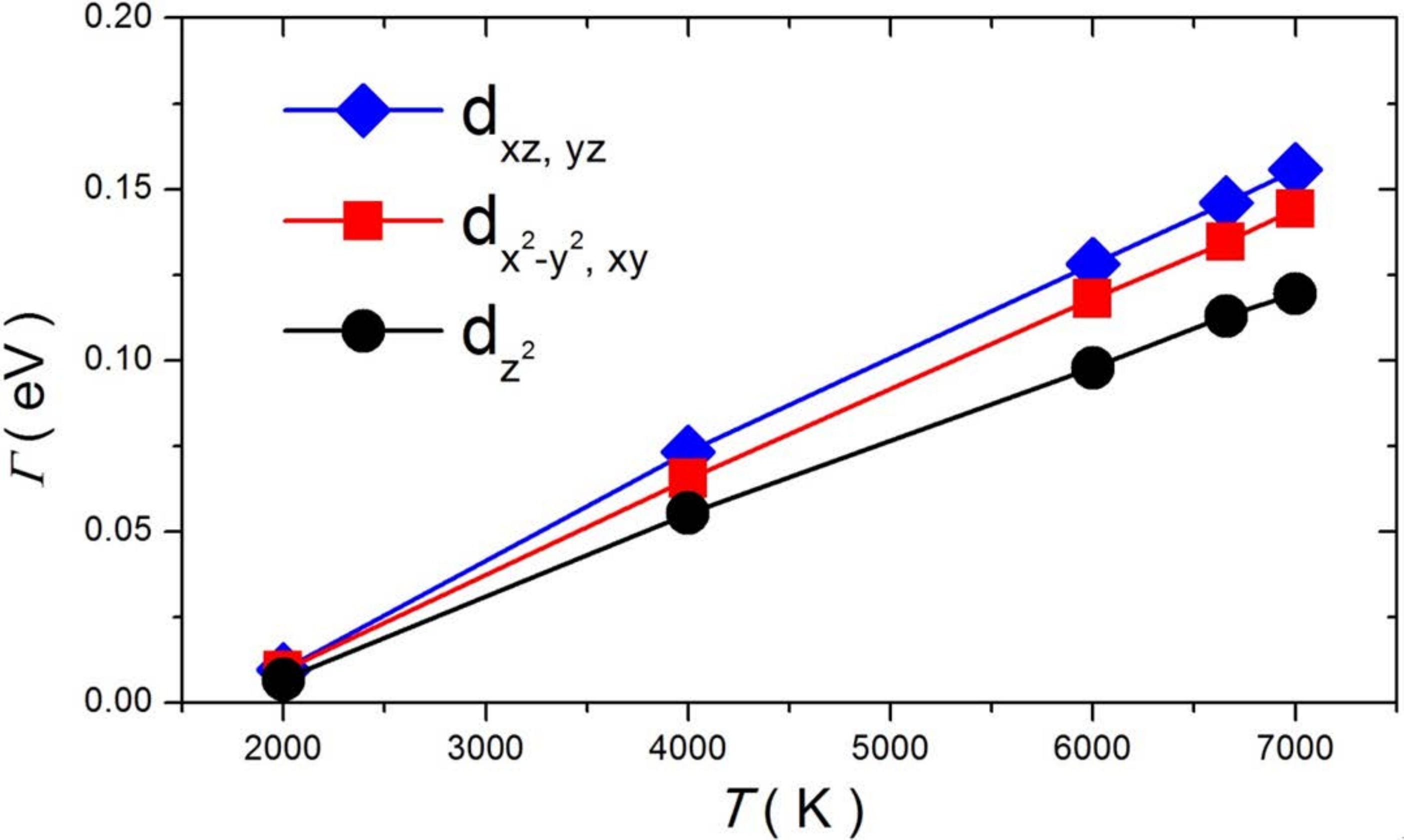}
  \caption{Orbitally resolved scattering rates of hcp due to electron-electron scattering at atomic volume 47.8 bohr$^3$.}
  \label{figs1}
\end{figure*}

\begin{figure*}
  \centering
  \noindent\includegraphics[width=1.0\columnwidth]{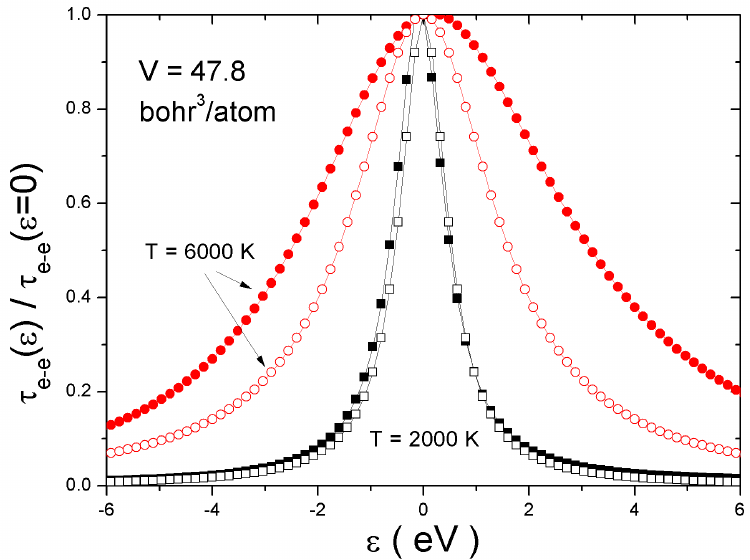}
  \caption{Energy-dependent part of the relaxation time due to electron-electron scattering. The filled squares  and circles are calculated from self-energy data given by Dynamical Mean Field Theory using $\tau(\varepsilon) = \hbar / ( - 2 Im\Sigma(\varepsilon) )$, at V = 47.8 bohr$^3$ / atom and T = 2000 K and 6000 K, respectively. The open squares and circles are for a Fermi liquid, which gives energy-dependent part $1 / [ 1 + \varepsilon^2 / ( \pi k_B T )^2 ]$, at the same conditions. The relaxation times at the chemical potential are 28.5 and 2.35 fs at T = 2000 K and 6000 K, respectively.}
  \label{figs2}
\end{figure*}

\begin{figure*}
  \centering
  \noindent\includegraphics[width=1.0\columnwidth]{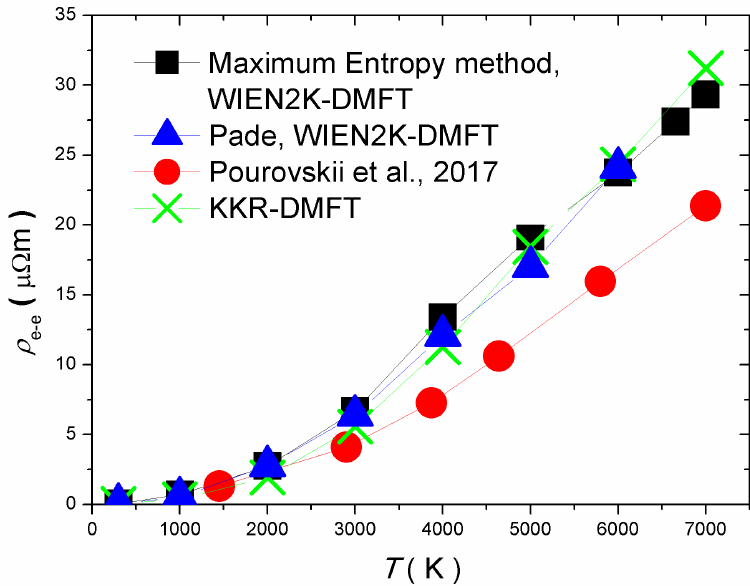}
  \caption{Our calculated $\rho_{e-e}$ compared with those calculated by $Pourovskii\ et\ al.$ \cite{pourovskii2017electron} at atomic volume 47.8 bohr$^3$.}
  \label{figs3}
\end{figure*}

\begin{figure*}
  \centering
  \noindent\includegraphics[width=1.0\columnwidth]{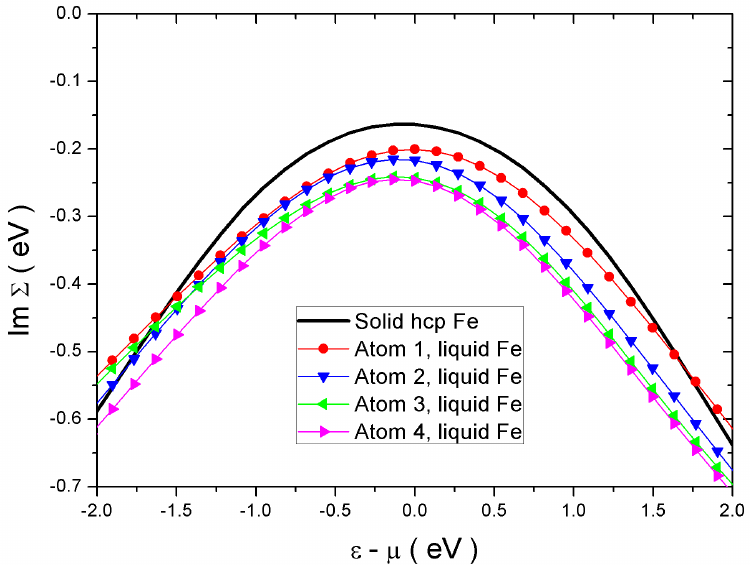}
  \caption{Im$\Sigma$ of 4 atoms of one snapshot of liquid Fe at 6.5 ps compared with Im$\Sigma$ of solid hcp Fe at atomic volume 56.5 bohr$^3$ and temperature of 4000 K.
  Im$\Sigma$ is averaged on ten d-orbitals.}
  \label{figs4}
\end{figure*}

\begin{table*}{
  \caption{\label{rho_2000} Electrical resistivity at about 100 GPa and 2000 K. Our DMFT+DFPT results used the experimental equation of state by $Dewaele\ et\ al.$ \cite{dewaele2006quasihydrostatic}. All FPMD results used their own theoretical equations of state. V is atomic volume.}
  \centering
  \begin{tabular}{cccccccc}
  \hline
  System & P(GPa) &
  V(bohr$^3$)
  & T(K) &
  $\rho_{e-ph}$($\mu\Omega$ m) & $\rho_{e-e}$($\mu\Omega$ m) & $\rho$($\mu\Omega$ m) &
  Type and Source \\
  \hline
  Solid Fe & 110 &  & 2180 & & & 53.5 &
  Shock Exp., $Keeler\ et\ al.$ \cite{keeler1969electrical}\\
  Solid Fe & 106 &  & 2000 & & & 31.1 &
  DAC Exp., $Ohta\ et\ al.$ \cite{ohta2016experimental}\\
  Solid Fe & 101 &  & 2010 & & & 37.6 &
  Model, $Gomi\ et\ al.$ \cite{gomi2013high}\\
  Liquid Fe & 122 & 55.9 & 2000 & 65.3 & & &
  FPMD, $de\ Koker\ et\ al.$ \cite{de2012electrical}\\
  Solid Fe & 73 & 60.4 & 2350 & 52.7 & & &
  FPMD, $de\ Koker\ et\ al.$ \cite{de2012electrical}\\
  Solid Fe & 102 & 57.9 & 2000 & 40.8 & 8.3 & 49.1 &
  DFPT+DMFT, this study \\
  \hline
  \end{tabular}
}\end{table*}

\begin{table*}{
  \caption{\label{kappa_2000} Thermal conductivity at about 100 GPa and 2000 K.}
  \centering
  \begin{tabular}{cccccccc}
  \hline
  System & P(GPa) & V(bohr$^3$) & T(K) & $\kappa_{e-ph}$(W/m/K) & $\kappa_{e-e}$(W/m/K) & $\kappa$(W/m/K) & Type and Source \\
  \hline
  Solid Fe & 112 & & 1700-2300 & & & 41.9-69.9 & DAC Exp., $Kon\hat{o}pkov\acute{a}\ et\ al.$ \cite{konopkova2016direct}\\
  Solid Fe & 112 &  & 2000 & & & 42.0 & Model, $Kon\hat{o}pkov\acute{a}\ et\ al.$ \cite{konopkova2016direct}\\
  Liquid Fe & 122 & 55.9 & 2000 & 82.7 & & & FPMD, $de\ Koker\ et\ al.$ \cite{de2012electrical}\\
  Solid Fe & 102 & 57.9 & 2000 & 111 & 362 & 84.7 & DFPT+DMFT, this study \\
  \hline
  \end{tabular}
}\end{table*}

\begin{table*}{
  \caption{\label{rho_cmb} Electrical resistivity at Core-Mantle Boundary conditions.}
  \centering
  \begin{tabular}{cccccccc}
  \hline
  System & P(GPa) & V(bohr$^3$) & T(K) & $\rho_{e-ph}$($\mu\Omega$ m) & $\rho_{e-e}$($\mu\Omega$ m) & $\rho$($\mu\Omega$ m) & Type and Source \\
  \hline
  Solid Fe & 140 &  & 2950 & & & 64.1 & Shock Exp., $Keeler\ et\ al.$ \cite{keeler1969electrical}\\
  Solid Fe & 146 &  & 2540 & & & 40.2 & DAC Exp., $Ohta\ et\ al.$ \cite{ohta2016experimental}\\
  Solid Fe & 135 &  & 3750 & & & 53.7 & Model, $Gomi\ et\ al.$ \cite{gomi2013high}\\
  Liquid Fe & 133 & 55.9 & 3000 & 65.4 & & & FPMD, $de\ Koker\ et\ al.$ \cite{de2012electrical}\\
  Liquid Fe & 151 & 55.9 & 4000 & 67.9 & & & FPMD, $de\ Koker\ et\ al.$ \cite{de2012electrical}\\
  Liquid Fe$_3$Si & 136 & 51.9 & 3000 & 102 & & & FPMD, $de\ Koker\ et\ al.$ \cite{de2012electrical}\\
  Liquid Fe & 124 & 59.5 & 4630 & 74.7 & & & FPMD, $Pozzo\ et\ al.$ \cite{pozzo2012thermal}\\
  Liquid Fe$_{79}$Si$_8$O$_{13}$ & 134 & 63.3 & 4112 & 90.5 & & & FPMD, $Pozzo\ et\ al.$ \cite{pozzo2012thermal}\\
  Solid Fe & 146 & 55.1 & 3360 & 55.6 & 16.4 & 72.0 & DFPT+DMFT, this study \\
  Solid Fe & 134 & 56.5 & 3750 & 61.6 & 23.7 & 85.3 & DFPT+DMFT, this study \\
  \hline
  \end{tabular}
}\end{table*}

\begin{table*}{
  \caption{\label{kappa_cmb} Thermal conductivity at Core-Mantle Boundary conditions.}
  \centering
  \begin{tabular}{cccccccc}
  \hline
  System & P(GPa) & V(bohr$^3$) & T(K) & $\kappa_{e-ph}$(W/m/K) & $\kappa_{e-e}$(W/m/K) & $\kappa$(W/m/K) & Type and Source \\
  \hline
  Solid Fe & 112 & & 2700-3000 & & & 19.8-34.6 & DAC Exp., $Kon\hat{o}pkov\acute{a}\ et\ al.$ \cite{konopkova2016direct}\\
  Solid Fe & 136 &  & 3800-4800 & & & 33$\pm$7 & Model, $Kon\hat{o}pkov\acute{a}\ et\ al.$ \cite{konopkova2016direct}\\
  Liquid Fe & 133 & 55.9 & 3000 & 110 & & & FPMD, $de\ Koker\ et\ al.$ \cite{de2012electrical}\\
  Liquid Fe & 151 & 55.9 & 4000 & 133 & & & FPMD, $de\ Koker\ et\ al.$ \cite{de2012electrical}\\
  Liquid Fe$_3$Si & 136 & 51.9 & 3000 & 75.3 & & & FPMD, $de\ Koker\ et\ al.$ \cite{de2012electrical}\\
  Liquid Fe & 124 & 59.5 & 4630 & 154 & & & FPMD, $Pozzo\ et\ al.$ \cite{pozzo2012thermal}\\
  Liquid Fe$_{79}$Si$_8$O$_{13}$ & 134 & 63.3 & 4112 & 99 & & & FPMD, $Pozzo\ et\ al.$ \cite{pozzo2012thermal}\\
  Solid Fe & 146 & 55.1 & 3360 & 146 & 397 & 98.0 & DFPT+DMFT, this study \\
  Solid Fe & 134 & 56.5 & 3750 & 130 & 340 & 93.8 & DFPT+DMFT, this study \\
  \hline
  \end{tabular}
}\end{table*}

\begin{table*}{
  \caption{\label{rho_icb} Electrical resistivity at Inner Core Boundary conditions.
  P L, $Pourovskii\ et\ al.$.}
  \centering
  \begin{tabular}{cccccccc}
  \hline
  System & P(GPa) & V(bohr$^3$) & T(K) & $\rho_{e-ph}$($\mu\Omega$ m) & $\rho_{e-e}$($\mu\Omega$ m) & $\rho$($\mu\Omega$ m) & Type and Source \\
  \hline
  Solid Fe & 330 &  & 4971 & & & 43.1 & Model, $Gomi\ et\ al.$ \cite{gomi2013high}\\
  Liquid Fe & 327 & 47.9 & 6000 & 62.0 & & & FPMD, $de\ Koker\ et\ al.$ \cite{de2012electrical}\\
  Fe$_3$Si & 318 & 47.9 & 6000 & 91.8 & & & FPMD, $de\ Koker\ et\ al.$ \cite{de2012electrical}\\
  Liquid Fe & 339 & 47.6 & 6420 & 64.1 & & & FPMD, $Pozzo\ et\ al.$ \cite{pozzo2012thermal}\\
  Liquid Fe$_{79}$Si$_8$O$_{13}$ & 328 & 52.0 & 5500 & 80.5 & & & FPMD, $Pozzo\ et\ al.$ \cite{pozzo2012thermal}\\
  Solid Fe & 365 & 46.2 & 6350 & 54.0 & & & FPMD, $Pozzo\ et\ al.$ \cite{pozzo2014thermal}\\
  Solid Fe$_{79}$Si$_8$O$_{13}$ & 329 & & 5500 & 64.3 & & & FPMD, $Pozzo\ et\ al.$ \cite{pozzo2014thermal}\\
  Solid Fe & 305 & 47.8 & 6000 & 63.2 & 23.8 & 87.0 & DFPT+DMFT, this study \\
  Solid Fe & & 47.8 & 6000 & & 16.0 & & DFPT+DMFT, P L \cite{pourovskii2017electron}\\
  \hline
  \end{tabular}
}\end{table*}

\begin{table*}{
  \caption{\label{kappa_icb} Thermal conductivity at Inner Core Boundary conditions.}
  \centering
  \begin{tabular}{cccccccc}
  \hline
  System & P(GPa) & V(bohr$^3$) & T(K) & $\kappa_{e-ph}$(W/m/K) & $\kappa_{e-e}$(W/m/K) & $\kappa$(W/m/K) & Type and Source \\
  \hline
  Solid Fe & 330 &  & 5600-6500 & & & 46$\pm$9 & Model, $Kon\hat{o}pkov\acute{a}\ et\ al.$ \cite{konopkova2016direct}\\
  Liquid Fe & 327 & 47.9 & 6000 & 215 & & & FPMD, $de\ Koker\ et\ al.$ \cite{de2012electrical}\\
  Liquid Fe$_3$Si & 318 & 47.9 & 6000 & 155 & & & FPMD, $de\ Koker\ et\ al.$ \cite{de2012electrical}\\
  Liquid Fe & 328 & 48.3 & 6350 & 246 & & & FPMD, $Pozzo\ et\ al.$ \cite{pozzo2012thermal}\\
  Liquid Fe$_{79}$Si$_8$O$_{13}$ & 328 & 52.0 & 5500 & 148 & & & FPMD, $Pozzo\ et\ al.$ \cite{pozzo2012thermal}\\
  Solid Fe & 329 & & 6350 & 313 & & & FPMD, $Pozzo\ et\ al.$ \cite{pozzo2014thermal}\\
  Solid Fe$_{79}$Si$_8$O$_{13}$ & 329 & & 5500 & 232 & & & FPMD, $Pozzo\ et\ al.$ \cite{pozzo2014thermal}\\
  Solid Fe & 305 & 47.8 & 6000 & 205 & 514 & 147 & DFPT+DMFT, this study \\
  Solid Fe & & 47.8 & 6000 & & 542 & & DFPT+DMFT, P L \cite{pourovskii2017electron}\\
  \hline
  \end{tabular}
}\end{table*}

\end{document}